\documentclass[letterpaper,titlepage,11pt]{article}

\usepackage{amssymb,amsmath,amsfonts}
\usepackage{epsfig}
\usepackage{ccaption}
\usepackage{graphicx}

\usepackage[
      colorlinks=black,
      linkcolor=black,
      urlcolor=black,
      filecolor=black,
      citecolor=black,
      pdfstartview=FitV,
      pdftitle={},
        pdfauthor={Kentaro Tanabe},
        pdfsubject={},
        pdfkeywords={},
        pdfpagemode=None,
        bookmarksopen=true
      ]{hyperref}

\def\clock{{\count0=\time
           \divide\count0 60
           \ifnum\count0<10 0\fi\the\count0
           \multiply\count0 -60 \advance\count0 \time
           :\ifnum\count0<10 0\fi \the\count0
         }}
\newcommand{\timestamp}{{\small\vbox{\hbox{\tt\jobname.tex}
\hbox{\the\day/\the\month/\the\year, \clock}}}}

\newcommand{\beq}{\begin{equation}}
\newcommand{\eeq}{\end{equation}}
\newcommand{\bea}{\begin{eqnarray}}
\newcommand{\eea}{\end{eqnarray}}
\newcommand{\beqa}{\begin{eqnarray}}
\newcommand{\eeqa}{\end{eqnarray}}

\newcommand{\sR}{\mathsf{R}}

\newcommand{\G}{\mathsf{G}_{D}}

\newcommand{\tH}{\text{H}}

\newcommand{\pz}{\partial_{z}}
\newcommand{\pv}{\partial_{v}}
\newcommand{\pp}{\partial_{\phi}}

\numberwithin{equation}{section}

\begin{document}

\begin{titlepage}
\rightline{KEK-TH-1867} 
\vskip 1.cm
\centerline{\LARGE \bf Black rings at large $D$} 
\vskip 1.cm
\centerline{\bf Kentaro Tanabe}
\vskip 0.5cm
\centerline{\sl Theory Center, Institute of Particles and Nuclear Studies, KEK,}
\centerline{\sl  Tsukuba, Ibaraki, 305-0801, Japan}
\vskip 0.5cm
\centerline{\small\tt ktanabe@post.kek.jp}

\vskip 1.cm
\centerline{\bf Abstract} \vskip 0.2cm 
\noindent 
We study the effective theory of slowly rotating black holes at the infinite limit of the spacetime dimension $D$. This large $D$ effective theory is obtained 
by integrating the Einstein equation with respect to the radial direction. The effective theory 
gives equations for non-linear dynamical deformations of a slowly rotating 
black hole by effective equations. 
The effective equations contain the slowly 
rotating Myers-Perry black hole, slowly boosted black string, non-uniform black string and black ring as stationary solutions.
We obtain the analytic solution of the black ring by solving effective equations. Furthermore, by perturbation analysis of effective equations, we find a quasinormal mode condition of the black ring in analytic way. As a result we confirm that 
thin black ring is unstable against non-axisymmetric perturbations. We also include $1/D$ corrections to the effective equations 
and discuss the effects by $1/D$ corrections.

\end{titlepage}
\pagestyle{empty}
\small
\tableofcontents
\normalsize
\newpage
\pagestyle{plain}
\setcounter{page}{1}

\section{Introduction}

Stationary asymptotically flat black holes have only spherical horizon topology in four dimensions by the Hawking's topology theorem \cite{Hawking:1971vc}. This 
situation changes drastically in higher dimensions, and higher dimensional black hole has various horizon topology. As one concrete example of this fact, Emparan and Reall found the five dimensional black hole solution, {\it black ring},  
whose horizon topology is $S^{1}\times S^{2}$ \cite{Emparan:2001wn}.  This is the first asymptotically flat black hole solution which has a non-spherical horizon topology. 
After this discovery many new five dimensional black hole solutions with multiple horizons have been constructed by the solution generating technique such as the inverse scattering method ( {\it e.g.,} see \cite{Emparan:2008eg} for a review). In these solutions the topology of each horizon is $S^{3}$ or $S^{1}\times S^{2}$. We expect that black hole solutions can have much more various and non-trivial horizon topology in higher 
dimensions than five. To study this 
variety it would be the first step to extend the Emparan-Reall's discovery to higher dimensions, that is, the construction of the black ring in higher dimensions. 
However the solution generating technique cannot be applied to the asymptotically flat solution in higher dimensions than five.  
Then there are some efforts to construct the black ring in higher dimensions by numerical methods \cite{Kleihaus:2012xh,Dias:2014cia} and by
analytical approximation methods such as the blackfold method \cite{Emparan:2007wm,Armas:2014bia}\footnote{
The blackfold method has achieved a great success along this direction. In fact black holes with various horizon topology, including expected and unexpected ones, has been found \cite{Emparan:2009vd,Armas:2015nea}. 
}. 

In this paper we consider the construction of the $D$ dimensional black ring solution by using another analytical approximation method, {\it the large $D$ expansion method} 
\cite{Asnin:2007rw, Emparan:2013moa}. 
Recently this large $D$ expansion method was extended to construct the stationary \cite{Emparan:2015hwa,Suzuki:2015iha} and time-dependent black hole solutions
\cite{Bhattacharyya:2015dva,Emparan:2015gva}. So it is interesting to see if we can obtain the black ring 
solution by the large $D$ expansion method. The paper \cite{Suzuki:2015iha} discussed the effective theory and gave the effective equation for
the large $D$ stationary rotating black holes. This stationary solution was assumed to have the $O(1)$ horizon  angular velocity at the large $D$ limit. On the other hand 
the analysis by the blackfold method \cite{Emparan:2007wm} found that the horizon angular velocity of the $D$ dimensional thin black ring becomes
%
\begin{eqnarray}
\Omega_{\tH}=\frac{1}{\sqrt{D-3}}\frac{1}{R},
\end{eqnarray}
%
where $R$ is a ring radius. This means that the horizon angular velocity of the thin black ring is $O(1/\sqrt{D})$, not $O(1)$ at the large $D$ limit. 
Thus the effective equation in \cite{Suzuki:2015iha}  cannot be straightforwardly applied to the (thin) black ring. Hence, to construct the black ring
solution by the large $D$ expansion method, we should study the effective theory of the black hole with $O(1/\sqrt{D})$ horizon angular velocity anew. 
The purposes of this paper are to give the effective theory of black holes with $O(1/\sqrt{D})$ horizon angular velocity by including 
time-dependence and to construct the large $D$ black ring solution by solving the effective equations. We call a black hole with $O(1/\sqrt{D})$ horizon angular velocity slowly rotating large $D$ black hole in this paper. We expect that there is no black ring with $O(1)$ horizon angular velocity. The horizon angular velocity of the black ring is determined by the balance condition between the string tension and centrifugal force by the boost, which is the horizon angular velocity. At large $D$ the tension of a black string becomes small compared to its mass by $O(D^{-1})$ \cite{ Emparan:2013moa}. As a result the horizon angular velocity should be also small at large $D$. Hence, generically, there is no black ring with $O(1)$ horizon angular velocity at large $D$. This is why we do not use the word ``slowly rotating black ring" in 
our analysis.   

The large $D$ effective theory for dynamical black holes has been considered in \cite{Bhattacharyya:2015dva,Emparan:2015gva}, where it was shown that the large $D$ expansion method gives 
simple effective equations for the dynamical black holes. The equations for dynamical black branes obtained in \cite{Emparan:2015gva} can be solved very easily, and the numerical solution
of the equations capture the non-linear evolution and endpoint of the Gregory-Laflamme instability of the black brane \cite{Gregory:1993vy} in higher dimensions than the critical dimension \cite{Sorkin:2004qq}. 
We will study the large $D$  effective equation for slowly rotating large $D$ black holes in the similar manner with \cite{Emparan:2015gva}. Then we
will see that the large $D$ black ring is found as an analytic stationary solution of the effective equations. Furthermore, by perturbing effective equations, 
we can obtain a quasinormal mode condition of the black ring. As found numerically in \cite{Santos:2015iua}, the quasinormal 
mode says that the thin black ring is unstable against non-axisymmetric perturbations, and relatively fat black ring is stable. Our large $D$ effective equations describe the non-linear evolution
of such non-axisymmetric instabilities of the black ring. We can also include the $1/D$ corrections to the effective equations. We find that the
$1/D$ corrections give striking features to the instability modes of the black ring. 

This paper is organized as following. In section \ref{2} we give the effective theory of slowly rotating large $D$ black holes by the effective equations. 
We also discuss some general properties of the stationary solution of the equations and give formula for thermodynamic quantities such as the mass and 
angular momentum of the solution with $1/D$ corrections. In section \ref{3} we construct the black ring solution analytically by solving the 
effective equations. Some physical properties such as quasinormal modes and phase diagram of the black ring will be also discussed there. We close
this paper by giving discussion and outlook of this work in section \ref{4}. The appendices contain technical details and some useful byproducts of the main results in this paper. Especially the slowly rotating Myers-Perry black hole and slowly boosted black string will be rediscovered  with their quasinormal mode frequencies analytically in Appendix \ref{B}. 
  
Note that we perform the large $D$ expansion of the Einstein equation to find $D$ dimensional black hole solutions. In the expansion, defining $n$ by
%
\begin{eqnarray}
D=n+4,
\end{eqnarray}
%
we use $1/n$ as the expansion parameter instead of $1/D$ in this paper. This definition of $n$ is same with one of \cite{Emparan:2007wm,Armas:2014bia} 
for black rings.

\section{Effective equations}\label{2}

We consider the large $D$ effective theory for slowly rotating large $D$ black holes. The effective equations will be given 
as equations for the energy and momentum density of a dynamical black hole. In this section we study general properties 
of solutions of the effective equations without specifying an embedding of solutions.  

\subsection{Setup}

The effective equations describe non-linear dynamical deformations of black holes. To obtain them, it is natural to use the ingoing 
Eddington-Finkelstein coordinate for the metric ansatz as
%
\begin{eqnarray}
&&
ds^{2} = - A dv^{2} -2(u_{v}dv+u_{a}dX^{a})dr \notag \\
&&~~~~~~~~~~~~~
+r^{2}G_{ab}dX^{a}dX^{b}
-2C_{a}dvdX^{a} +r^{2}H^{2} d\Omega^{2}_{n},
\label{mansatz}
\end{eqnarray}
%
where $X^{a}=(z,\Phi)$. 
$A$, $u_{v}$, $G_{ab}$, $C_{a}$ 
are functions of $(v,r,X^{a})$. $u_{a}$ and $H$ are functions of $(v,X^{a})$. In this paper we use $v$ as a time coordinate in this 
system and the effective theory obtained below, while $t$ was used for it in \cite{Emparan:2015gva}.
This metric ansatz partially fixes the gauge of the radial coordinate $r$. The residual gauge can be fixed
in solving the Einstein equations. $X^{a}=(z,\Phi)$ are inhomogeneous spatial directions of black holes, and we take $\Phi$ as the rotational direction. 
We give large $D$ scalings and boundary conditions of the metric functions. 

\paragraph{Large $D$ scalings}
To solve the Einstein equation by $1/n$ expansion, we should specify 
behaviors of each metric functions at large $D$. In the analysis of the blackfold \cite{Emparan:2007wm} they found that 
the horizon angular velocity of the $D=n+4$ dimensional thin black ring is given by
%
\begin{eqnarray}
\Omega_{H} = \frac{1}{\sqrt{n+1}}\frac{1}{R}+O(R^{-2}),
\label{BFformula}
\end{eqnarray}
%
where $R$ is a ring radius. This result implies that the metric function $C_{\Phi}$ in eq. (\ref{mansatz}) becomes $O(1/\sqrt{n})$ at large $D$. 
From this observation we define the slowly rotating large $D$ black hole by $C_{\Phi}=O(1/\sqrt{n})$
\footnote{
The stationary solution considered in \cite{Suzuki:2015iha} was assumed to have $C_{\Phi}=O(1)$. 
}.  
One may think that this assumption is restrictive for a black ring and describes the dynamics only of a thin black ring solution.   
However this is not correct. As we will see, solutions obtained in this paper cover wider range than one by the blackfold because we do not require the much 
larger ring radius than the ring thickness. In our analysis ring radius and thickness can be comparable, and our large $D$ solution can describe 
the dynamics also of not-thin black ring. 

In our setting $\partial_{\Phi}$ is not a Killing vector. Thus the metric functions have $\Phi$-dependences in general. Then, to have 
consistent $1/n$ expansions of the Einstein equations, we should assume $\partial_{\Phi}=O(\sqrt{n})$
\footnote{
We consider the decoupled mode excitation \cite{Emparan:2014aba}. So we assume $\pv=O(1)$ and $\partial_{r}=O(D)$. 
}. If we take $\partial_{\Phi}=O(1)$, the Einstein equations have not only $1/n$ series but also $1/\sqrt{n}$ series, and 
the analysis becomes a bit involved. To treat the scaling $\partial_{\Phi}=O(\sqrt{n})$ it is useful to introduce new coordinate 
$\phi$ defined by
%
\begin{eqnarray}
\Phi = \frac{\phi}{\sqrt{n}}. 
\label{phiRe}
\end{eqnarray}
%
Then $\partial_{\Phi}=O(\sqrt{n})$ is equivalent to $\pp=O(1)$. 
Summarizing, our metric ansatz for the slowly rotating large $D$ black holes is
%
\begin{eqnarray}
&&
ds^{2} = - A dv^{2} -2(u_{v}dv+u_{a}dx^{a})dr \notag \\
&&~~~~~~~~~~~~~~~~~
+r^{2}G_{ab}dx^{a}dx^{b}
-2C_{a}dvdx^{a} +r^{2}H^{2} d\Omega^{2}_{n},
\label{metric}
\end{eqnarray}
%
where $x^{a}=(z,\phi)$. Each metric functions have following large $D$ expansions
%
\begin{eqnarray}
A = \sum_{k\geq 0} \frac{A^{(k)}(v,r,x^{a})}{n^{k}},~~
C_{z} = \sum_{k\geq 0} \frac{C_{z}^{(k)}(v,r,x^{a})}{n^{k+1}},~~
C_{\phi} = \sum_{k\geq 0} \frac{C_{\phi}^{(k)}(v,r,x^{a})}{n^{k+1}},
\end{eqnarray}
%
%
\begin{eqnarray}
u_{v} = \sum_{k\geq 0} \frac{u_{v}^{(k)}(v,r,x^{a})}{n^{k}},~~
u_{z} = \sum_{k\geq 0} \frac{u_{z}^{(k)}(v,x^{a})}{n^{k+1}},~~
u_{\phi} = \sum_{k\geq 0} \frac{u_{\phi}^{(k)}(v,x^{a})}{n^{k+1}},
\end{eqnarray}
%
%
\begin{eqnarray}
G_{zz} = 1+ \sum_{k\geq 0} \frac{G_{zz}^{(k)}(v,r,x^{a})}{n^{k+1}},~~
G_{z\phi} = \sum_{k\geq 0} \frac{G_{z\phi}^{(k)}(v,r,x^{a})}{n^{k+2}},~~
\end{eqnarray}
%
and
%
\begin{eqnarray}
G_{\phi\phi} = \frac{G(z)^{2}}{n}
\left( 1+\sum_{k\geq 0} \frac{G_{\phi\phi}^{(k)}(v,r,x^{a})}{n^{k+1}} \right),~~
H=H(z)
\label{GHansatz}
\end{eqnarray}
%
Under these scaling assumptions, the Einstein equation can be consistently expanded in $1/n$. 
$\phi$ is the rotational direction of the black hole, and we assume $\pv$ and $\pp$ become the Killing vectors in a background geometry.
This assumption implies that the asymptotic metric at far from the horizon does not have $v$ and $\phi$ dependences. So the metric function $H(z)$ and the leading order of $G_{\phi\phi}$
has only $z$ dependences as seen in eq. (\ref{GHansatz}). Finally we introduce new radial coordinate $\sR$ to realize $\partial_{r}=O(n)$ defined by
%
\begin{eqnarray}
\sR = \left( \frac{r}{r_{0}} \right)^{n},
\end{eqnarray}
%
where a constant $r_{0}$ is a fiducial horizon size, and we set to $r_{0}=1$. 

\paragraph{Boundary conditions}

We give boundary conditions for metric functions. In this paper we consider non-linear dynamics of the decoupled mode excitations \cite{Emparan:2014aba}. 
The decoupled mode condition demands following boundary conditions in the asymptotic region $\sR\gg 1$ 
%
\begin{eqnarray}
A=1+O(\sR^{-1}),~~
C_{a} = O(\sR^{-1})~~,
G_{z\phi}=O(\sR^{-1}).
\end{eqnarray}
%
The dynamical solutions have the horizon-like surface where $A$ vanishes at the leading order in $1/n$ expansion
\footnote{
In this paper we consider the slowly rotating black holes. Thus the horizon position is determined mainly by $g_{uu}=-A$. 
}. We regard the surface as a horizon of the dynamical black hole. The boundary condition on the horizon is the regularity condition
of the each metric functions.  

\subsection{Effective equations}
We solve the Einstein equation by the large $D$ expansion. The leading order equations of the Einstein equations contain only $\sR$-derivatives, 
so we can integrate them easily. Then, the leading order solutions after imposing the boundary conditions are obtained as
%
\begin{eqnarray}
A^{(0)}=1-\frac{p_{v}(v,x^{a})}{\sR},~~
C_{a}^{(0)}=\frac{p_{a}(v,x^{a})}{\sR},~~
u_{v}^{(0)} = -\frac{H(z)}{\sqrt{1-H'(z)^{2}}},
\end{eqnarray}
%
%
\begin{eqnarray}
u_{a}^{(0)}=u_{a}^{(0)}(z),~~
G_{zz}^{(0)}=0,~~
G_{z\phi}^{(0)}=\frac{p_{z}p_{\phi}}{p_{v}}\frac{1}{\sR},
\label{LOsolGzx}
\end{eqnarray}
%
and
%
\begin{eqnarray}
G_{\phi\phi}^{(0)}= -\left( 2+ \frac{2H(z)G'(z)H'(z)}{G(z)(1-H'(z)^{2})}\right)\log{\sR} +\frac{p_{\phi}^{2}}{G(z)^{2}p_{v}}\frac{1}{\sR}.
\end{eqnarray}
%
$p_{v}(v,x^{a})$ and $p_{a}(v,x^{a})$ are mass and momentum density of the solution. 
They are introduced as integration functions of $\sR$-integrations of the Einstein 
equations. We can see that the horizon position of this dynamical solution 
is at $\sR=p_{v}(v,x^{a})$. Furthermore we found that the function $H(z)$ should satisfy the following condition 
%
\begin{eqnarray}
1-H'(z)^{2}+H(z)H''(z)=0,
\label{HcondLO}
\end{eqnarray}
%
from the boundary condition of $G_{z\phi}^{(0)}$ at asymptotic region $\sR\gg 1$. In eq. (\ref{LOsolGzx}) we already imposed the condition (\ref{HcondLO}). 
This condition is equivalent to the constant condition of the mean curvature of 
$r=\text{const.}$ surface in the asymptotic region for static solutions \cite{Emparan:2015hwa}\footnote{
If we perform $(D-1)+1$ decomposition of the Einstein equation on $\sR=\text{const.}$ surface, this condition is obtained from the momentum constraint
as shown in \cite{Emparan:2015hwa}. 
}. In this paper we consider 
small rotation $\Omega_{\tH}=O(1/\sqrt{n})$, so the leading order solution has no effects of rotations and gives same equations with static solutions. The condition (\ref{HcondLO}) 
is integrated as
%
\begin{eqnarray}
\frac{\sqrt{1-H'(z)^{2}}}{H(z)}= 2\hat{\kappa},
\label{Hcond}
\end{eqnarray}
%
where $\hat{\kappa}$ is a constant related with the surface gravity of the horizon as we will see later. So we call $\hat{\kappa}$ a 
reduced surface gravity. At the leading order the integration
functions $p_{v}$ and $p_{a}$ are arbitrary functions of $(v,x^{a})$, and there is no constraint among them. $u_{a}^{(0)}$ is 
also arbitrary function of $z$ at the leading order. At higher order in $1/n$, constraints for $u_{a}^{(0)}$ appear, 
and we can determine their explicit forms. 

Solving the next-to leading order Einstein equations we obtain non-trivial conditions which $p_{v}(v,x^{a})$ and $p_{a}(v,x^{a})$ should satisfy. 
The non-trivial conditions can be obtained also from the momentum constraints on $\sR=\text{const}.$ surface. 
These conditions give effective equations for the slowly rotating large $D$ black holes. Their forms are
%
\begin{eqnarray}
\pv p_{v} -\frac{H'(z)}{2\hat{\kappa}H(z)}\pz p_{v} -\frac{\pp^{2} p_{v}}{2\hat{\kappa}G(z)^{2}} 
+\frac{\pp p_{\phi}}{G(z)^{2}}+\frac{H'(z)}{H(z)}p_{z}=0,
\label{Deq1}
\end{eqnarray}
%
%
\begin{eqnarray}
&&
\pv p_{\phi} -\frac{H'(z)}{2\hat{\kappa}H(z)}\pz p_{\phi} -\frac{\pp^{2}p_{\phi}}{2\hat{\kappa}G(z)^{2}} 
+\frac{1}{G(z)^{2}}\pp \Biggl[ \frac{p_{\phi}^{2}}{p_{v}} \Biggr] \notag \\
&&~~~~~
-\frac{4\hat{\kappa}^{2}G(z)H(z)^{2}+2G'(z)H(z)H'(z)}{4\hat{\kappa}^{2}G(z)H(z)^{2}}\pp p_{v}
 \notag \\
&&~~~~~
+\frac{H'(z)}{H(z)}\frac{p_{z}p_{\phi}}{p_{v}}
+\frac{G'(z)H'(z)}{\hat{\kappa}G(z)H(z)}p_{\phi}=0,
\label{Deq2}
\end{eqnarray}
%
and
%
\begin{eqnarray}
&&
\pv p_{z} -\frac{H'(z)}{2\hat{\kappa}H(z)}\pz p_{z} -\frac{\pp^{2}p_{z}}{2\hat{\kappa}G(z)^{2}}
+\pz p_{v} +\frac{1}{G(z)^{2}}\pp \Biggl[ \frac{p_{\phi}p_{z}}{p_{v}} \Biggr] \notag \\
&&~~~~~~~
+\frac{H'(z)}{H(z)}\frac{p^{2}_{z}}{p_{v}}
-\frac{G'(z)}{G(z)^{3}}\frac{p^{2}_{\phi}}{p_{v}} +\frac{G'(z)}{\hat{\kappa}G(z)^{3}}\pp p_{\phi} \notag \\
&&~~~~~~~
+\frac{H(z)G'(z)^{2}H'(z)+G(z)(G'(z)-H(z)H'(z)G''(z))}{4\hat{\kappa}^{2}G(z)^{2}H(z)^{2}}p_{v} \notag \\
&&~~~~~~~
-\frac{1-2H'(z)^{2}}{2\hat{\kappa}H(z)^{2}}p_{z}=0.
\label{Deq3}
\end{eqnarray}
%
These equations describe dynamical non-linear deformations of mass and momentum density of the dynamical black hole. 
To solve these equations we should specify the embedding functions $G(z)$ and $H(z)$. Before doing that, we study some
general properties of solutions of these effective equations. Note that the effective equations for dynamical black strings 
obtained in \cite{Emparan:2015gva} can be reproduced from eqs. (\ref{Deq1}), (\ref{Deq2}) and (\ref{Deq3}) as one example. 

As one simple solution of the effective equations, we consider a stationary
solution
%
\begin{eqnarray}
p_{v}=p_{v}(z),~~p_{a}=p_{a}(z).
\end{eqnarray}
%
This solution has two Killing vectors $\pv$ and $\pp$. The effective equations can be solved under this ansatz. 
Eqs. (\ref{Deq1}) and (\ref{Deq2}) give  
%
\begin{eqnarray}
p_{z} = \frac{p_{v}'(z)}{2\hat{\kappa}},~~
p_{\phi}(z) = \hat{\Omega}_{H} G(z)^{2}p_{v}(z). 
\label{psol}
\end{eqnarray}
%
$\hat{\Omega}_{H}$ is an integration constant of $z$-integration of eq. (\ref{Deq2}). To see the physical meaning of 
$\hat{\Omega}_{H}$, we rewrite the $(v,\phi)$ part of the leading order metric of this stationary solution as
%
\begin{eqnarray}
ds^{2}_{(v,\phi)} = -\left( 1-\frac{p_{v}(z)}{\sR} \right) dv^{2}
+\frac{G(z)^{2}}{n}\left( d\phi - \hat{\Omega}_{H}\frac{p_{v}(z)}{\sR}dv \right)^{2} +O(1/n). 
\end{eqnarray}
%
From this expression we can see that $\Omega_{\tH}=\hat{\Omega}_{H}/\sqrt{n}$ gives the horizon angular velocity. Actually horizon generating Killing vector $\xi$ is 
%
\begin{eqnarray}
\xi =\frac{\partial}{\partial v} +\hat{\Omega}_{H}\frac{\partial}{\partial\phi}
=\frac{\partial}{\partial v} +\Omega_{H}\frac{\partial}{\partial\Phi}.
\end{eqnarray}
%
The Killing vector $\xi$ becomes null at the horizon $\sR=p_{v}(z)$. 
Furthermore we can calculate the surface gravity $\kappa$ of the black hole by using $\xi$ as
%
\begin{eqnarray}
\kappa &=& -\frac{\partial_{r}(\xi_{\mu}\xi^{\mu})}{2\xi_{r}}\Bigl|_{\sR=p_{v}(z)} \notag \\
&=& n\frac{\sqrt{1-H'(z)^{2}}}{2H(z)} \notag \\ 
&=& n \hat{\kappa},
\end{eqnarray}
%
where $x^{\mu}=(v,x^{a})$. Here we omit $O(1/n)$ terms for simplification. Thus the integration constant $\hat{\kappa}$ represents
the surface gravity of the black hole at the leading order in $1/n$ expansions. 
Finally, substituting the solutions (\ref{psol}) into eq. (\ref{Deq3}), we obtain an equation for $p_{v}(z)=e^{P(z)}$ as
%
\begin{eqnarray}
&&
P''(z) -\frac{H'(z)}{H(z)}P(z) \notag \\
&&
-\Bigl[
\frac{G'(z)^{2}}{G(z)^{2}}
-\frac{G''(z)}{G(z)}+\frac{G'(z)}{G(z)H(z)H'(z)}
-\hat{\Omega}^{2}_{\tH}\frac{G(z)G'(z)(1-H'(z)^{2})}{H(z)H'(z)}
\Bigr]=0.
\label{PeqS}
\end{eqnarray}
%
To solve this equation for $P(z)$ we should specify the functions $G(z)$ and $H(z)$. The functions can be determined by
embedding the solution into a background geometry. In next section we will embed the leading order solution into a flat 
background in the ring coordinate. Such solution describes dynamical black rings. In Appendix \ref{B} we show other 
solutions such as the Myers-Perry black hole and boosted black strings by considering the embeddings into 
other backgrounds.

\subsection{Effective energy-momentum tensor}

Next we study physical quantities of the solution. The mass and angular momentum of dynamical black holes can be
evaluated by the effective energy-momentum tensor. The effective energy momentum tensor,  $T_{\mu\nu}$, is defined by
\footnote{
The effective energy momentum tensor is originally defined on $r=\text{const}.$ surface. Thus its component runs over $(v,x^{a},x^{I})$ where
$x^{I}$ is a coordinate on $S^{n}$ in eq. (\ref{metric}). 
However the metric components of $g_{IJ}$ have been taken as non-dynamical ones by the gauge choice. So we consider only $x^{\mu}=(v,x^{a})$
components of the effective energy momentum tensor. 
}
%
\begin{eqnarray}
T_{\mu\nu} =-\frac{1}{8\pi \G} \left( \Bigl[ K_{\mu\nu}\Bigr]_{\Sigma}-\bar{g}_{\mu\nu}\Bigl[ K\Bigr]_{\Sigma} \right),
\end{eqnarray}
%
where $x^{\mu}=(v,x^{a})$ in eq. (\ref{metric}). $\Sigma$ is a $r=\text{const.}$ surface in the asymptotic region $\sR \gg 1$. $\G$ is the gravitational constant in $D$ dimensions. 
$K_{\mu\nu}$ is the extrinsic curvature of $r=\text{const.}$ surface. The square bracket represents the background 
subtraction at $\Sigma$. $\bar{g}_{\mu\nu}$ is the background induced metric on $\Sigma$.
The background metric $\bar{g}_{\mu\nu}$ in the asymptotic region $\sR\gg 1$ is obtained from the leading order solution as
%
\begin{eqnarray}
\bar{g}_{\mu\nu}dx^{\mu}dx^{\nu} &=& -dv^{2}+r^{2}dz^{2} \notag \\
&&
+\frac{G(z)^{2}}{n}\left( 1 -\frac{2H(z)H'(z)G'(z)\log{\sR}}{nG(z)(1-H'(z)^{2})} +O(n^{-2}) \right)d\phi^{2}.
\end{eqnarray}
%
The background extrinsic curvature, say $\bar{K}_{\mu\nu}$, is calculated as
%
\begin{eqnarray}
\bar{K}_{vv} = O(1/n),~~
\bar{K}_{zz} = 2\hat{\kappa} +O(1/n),~~
\bar{K}_{\phi\phi} = -\frac{G(z)G'(z)H'(z)}{2\hat{\kappa}H(z)}+O(1/n),
\end{eqnarray}
%
in our gauge choice for the radial coordinate $r$.
The trace part of the background extrinsic curvature, $\bar{K}$, contains contributions from $\bar{K}_{IJ}$ where
$x^{I}$ is a coordinate on $S^{n}$ in eq. (\ref{metric}). 
It is obtained as
%
\begin{eqnarray}
\bar{K} = n \Biggl[ 2\hat{\kappa}  
+\frac{1}{n}\left( 2\hat{\kappa} -\frac{G'(z)H'(z)}{2\hat{\kappa}G(z)H(z)} -2\hat{\kappa}\log{\sR} \right)
+O(n^{-2})\Biggr].  
\end{eqnarray}
%
By definition,  the background extrinsic curvature $\bar{K}_{\mu\nu}$ satisfies
%
\begin{eqnarray}
K_{\mu\nu} -\bar{K}_{\mu\nu} = O(1/\sR),~~
K -\bar{K} = O(1/\sR).
\end{eqnarray}
%
Then we can compute the effective energy-momentum tensor $T_{\mu\nu}$. The results are
%
\begin{eqnarray}
&&
T_{vv} = \frac{n\hat{\kappa}p_{v}}{8\pi \G\sR}\left( 1+O(n^{-1},\sR^{-1}) \right),~~
T_{va} = \frac{2\hat{\kappa}p_{a}-\partial_{a}p_{v}}{16\pi \G\sR}\left( 1+O(n^{-1},\sR^{-1}) \right),
\end{eqnarray}
%
and
%
\begin{eqnarray}
&&
T_{zz}=-\frac{G(z)^{2}(H'(z)p_{z}+H(z)\pz p_{v})+H(z)\pp p_{\phi}}{16\pi \G G(z)^{2}H(z)\sR}\left( 1+O(n^{-1},\sR^{-1}) \right), \notag \\
&&
T_{z\phi}=-\frac{1}{n}\frac{G(z)(2\hat{\kappa}p_{z}p_{\phi}-p_{v}(\pz p_{\phi}+\pp p_{z}))+2G'(z)p_{\phi}p_{v}}{16\pi \G G(z)p_{v} \sR}\left( 1+O(n^{-1},\sR^{-1}) \right), \notag \\
&&
T_{\phi\phi}=\frac{1}{n}\frac{1}{32\pi \G\hat{\kappa}H(z)p_{v} \sR}
\Bigl[
2\hat{\kappa}H(z)(2\hat{\kappa}p_{\phi}^{2}+p_{v}(\pp p_{\phi}+G(z)^{2}(\pv p_{v}-2\hat{\kappa}p_{v}))) \notag \\
&&~~~~~~~~~~~~~~~~
-G(z)H'(z)p_{v}(G'(z)p_{v}-2\hat{\kappa}G(z)p_{z})
\Bigr]\left( 1+O(n^{-1},\sR^{-1}) \right).
\end{eqnarray}
%
Then we can define the mass $\mathcal{M}$ and angular momentum $\mathcal{J}_{\Phi}$ of a dynamical black hole from this $T_{\mu\nu}$ as
%
\begin{eqnarray}
&&
\mathcal{M}=\frac{n\hat{\kappa}\Omega_{n}}{8\pi \G}\int dz\frac{d\phi}{\sqrt{n}} ~ G(z)H(z)^{n}p_{v}(v,x^{a}), 
\label{MLO}\\
&&
\mathcal{J}_{\Phi}=\sqrt{n}\frac{\hat{\kappa}\Omega_{n}}{8\pi \G}\int dz\frac{d\phi}{\sqrt{n}} ~ G(z)H(z)^{n}p_{\phi}(v,x^{a}),
\label{JLO}
\end{eqnarray}
%
where $H(z)^{n}$ term comes from the volume factor on $S^{n}$. $\Omega_{n}$ is the volume of unit $S^{n}$.
The $\sqrt{n}$ factor in $\mathcal{J}_{\Phi}$ comes from 
the relation between $\phi$ and $\Phi$ in eq. (\ref{phiRe}). For stationary solutions we can define the mass and angular momentum
of the black hole by the Komar integral of each Killing vector. It can be shown easily that their definitions are equivalent. 

The horizon area of the solution, $\mathcal{A}_{\tH}$, becomes
%
\begin{eqnarray}
\mathcal{A}_{\tH} = \Omega_{n}\int dz\frac{d\phi}{\sqrt{n}} ~ G(z)H(z)^{n}p_{v}(v,x^{a}), 
\label{AreaLO}
\end{eqnarray}
%
where we used the fact that the horizon is at $\sR=p_{v}$. 
Then we find that the stationary solution satisfies the Smarr formula given by
%
\begin{eqnarray}
\frac{n+1}{n+2}\mathcal{M}  =\frac{\kappa}{8\pi\G} \mathcal{A}+\Omega_{\tH} \mathcal{J}_{\Phi}, 
\label{Smarr}
\end{eqnarray}
%
at the leading order in $1/n$ expansions. Note that since $\Omega_{\tH}=O(1/\sqrt{n})$ and $\kappa=O(n)$, 
$\Omega_{\tH}\mathcal{J}_{\Phi}$ term does not contribute to 
the Smarr formula at the leading order. The mass and area of the solution can be calculated also for non-stationary solutions $p_{v}=p_v(v,x^{a})$. 
For this non-stationary solution, we can see that the Smarr formula (\ref{Smarr}) holds at the leading order in $1/n$ expansion. But this formula does not
hold at the higher order for non-stationary solutions.  

\subsection{$1/D$ corrections}

Solving the next-to-next-to leading order of the Einstein equations, we can obtain $1/n$ corrections to the effective equations (\ref{Deq1}), 
(\ref{Deq2}) and (\ref{Deq3}). The explicit forms of the $1/n$ corrections to the effective equations are not simple, so we show them in Appendix \ref{C}. 
Here we show only physical effects by $1/n$ corrections to the solution. 

At first we define the mass and momentum density, $p_{v}$ and $p_{a}$, up to $1/n$ corrections. The mass and momentum density are introduced as
integration functions of $\sR-$integrations of the Einstein equations. Thus we should specify the normalization of the integration functions. 
Expanding $A$ and $C_{a}$ up to $1/n$ corrections we define the mass and momentum density, $p_{v}$ and $p_{a}$, by coefficients of $1/\sR$ as
%
\begin{eqnarray}
A = 1 -\frac{p_{v}(v,x^{a})}{\sR} +O(n^{-2},\sR^{-2}),~~C_{a} = \frac{p_{a}(v,x^{a})}{\sR} +O(n^{-2},\sR^{-2}).
\end{eqnarray}
%
This definition normalizes the integration functions up to $O(1/n)$ by the asymptotic behavior of the metric functions. 
Using this definition and the next-to-next-to leading order stationary solutions, the surface gravity up to $1/n$ correction is obtained as
%
\begin{eqnarray}
\frac{\kappa}{n} = \hat{\kappa} -\frac{1}{n}\Biggl[
\frac{G'(z)H'(z)}{4\hat{\kappa}G(z)H(z)} +\hat{\kappa}\log{p_{v}} +\frac{H'(z)}{2H(z)}\frac{p_{z}}{p_{v}} 
+\frac{\hat{\kappa}}{2G(z)^{2}} \frac{p_{\phi}^{2}}{p_{v}^{2}}
\Biggr]+O(n^{-2}).
\end{eqnarray}
%
One may think that this expression is strange since the surface gravity $\kappa$ seems not to be constant when
the solution is stationary where $p_{v}=p_{v}(z)$ and $p_{a}=p_{a}(z)$. However we can show that 
%
\begin{eqnarray}
\frac{d}{dz}\Biggl[
\frac{G'(z)H'(z)}{4\hat{\kappa}G(z)H(z)} +\hat{\kappa}\log{p_{v}} +\frac{H'(z)}{2H(z)}\frac{p_{z}}{p_{v}}
+\frac{\hat{\kappa}}{2G(z)^{2}} \frac{p_{\phi}^{2}}{p_{v}^{2}}
\Biggr] =0,
\end{eqnarray}
%
by using the effective equations (\ref{Deq1}), (\ref{Deq2}) and (\ref{Deq3}) for the stationary solution. Hence the surface gravity becomes 
constant for the stationary solution as expected. The horizon angular velocity $\Omega_{\tH}=\hat{\Omega}_{\tH}/\sqrt{n}$ is also obtained as
%
\begin{eqnarray}
\hat{\Omega}_{\tH} = \frac{p_{\phi}}{G(z)^{2}p_{v}} 
+\frac{1}{n} \frac{G'(z)H'(z)}{2\hat{\kappa}^{2}G(z)^{3}H(z)}\frac{p_{\phi}\log{p_{v}}}{p_{v}}+O(n^{-2}).
\end{eqnarray}
%
The first term in r.h.s. is equivalent to $\hat{\Omega}_{\tH}$ at the leading order as seen in eq. (\ref{psol}). But, at higher order in $1/n$ expansion, they are different. 
We can show that, when the solution is stationary, the horizon angular velocity is constant up to $1/n$ correction by using the effective equations with $1/n$ corrections 
given in Appendix \ref{C}. 

We can calculate the mass, angular momentum and area of the solution up to $O(1/n)$. The results are complicate, so we show the results only for the stationary 
solutions $p_{v}=p_{v}(z)$ and $p_{a}=p_{a}(z)$. 
The mass and angular momentum formula up to $O(1/n)$ become 
%
\begin{eqnarray}
&&
\mathcal{M}=\frac{n\hat{\kappa}\Omega_{n}}{8\pi \G}\int dz\frac{d\phi}{\sqrt{n}} ~ G(z)H(z)^{n}M(z), \\
&&
\mathcal{J}_{\Phi}=\sqrt{n}\frac{\hat{\kappa}\Omega_{n}}{8\pi \G}\int dz\frac{d\phi}{\sqrt{n}} ~ G(z)H(z)^{n}J_{\phi}(z),
\label{MJdefNLO}
\end{eqnarray}
%
where
%
\begin{eqnarray}
&&
M(z) = p_{v} +\frac{1}{n}\Biggl[p_{v}
-\frac{p_{z}}{2\hat{\kappa}}\frac{H'(z)}{H(z)} -\frac{p_{v}}{4\hat{\kappa}^{2}}\frac{G'(z)H'(z)}{G(z)H(z)}
\Biggr],\label{MdefNLO}\\
&&
J_{\phi}(z) = p_{\phi} -\frac{1}{n}\Biggl[ \frac{p_{z}p_{\phi}}{2\hat{\kappa}p_{v}}\frac{H'(z)}{H(z)} 
+\frac{p_{\phi}}{4\hat{\kappa}^{2}}\frac{G'(z)H'(z)}{G(z)H(z)}
\Biggr]. \label{JdefNLO}
\end{eqnarray}
%
The mass and angular momentum can be obtained from the Komar integral for the stationary solution or from 
the effective energy momentum tensor for general time-dependent solution. These definitions coincide for the stationary solution.  
The formula of the horizon area also has the $1/n$ correction as
%
\begin{eqnarray}
\mathcal{A}_{\tH}=\Omega_{n}\int dz\frac{d\phi}{\sqrt{n}} ~ G(z)H(z)^{n}A_{\tH}(v,x^{a})
\label{areaNLO}
\end{eqnarray}
%
where
%
\begin{eqnarray}
A_{\tH} = p_{v}+\frac{1}{n}\Biggl[
p_{v}\log{p_{v}}-\frac{1}{2G(z)^{2}}\frac{p_{\phi}^{2}}{p_{v}}
\Biggr].
\label{areaNLO2}
\end{eqnarray}
%
Then we can see that the Smarr formula, 
%
\begin{eqnarray}
\frac{n+1}{n+2}\mathcal{M} = \frac{\kappa}{8\pi \G}\mathcal{A}_{\tH}+\Omega_{\tH}\mathcal{J}_{\Phi}, 
\end{eqnarray} 
%
can be satisfied up to $1/n$ corrections.

\section{Black ring and its physical properties}\label{3}

Next we solve the effective equations in an explicit embedding. Especially we consider an embedding into a flat background 
in the ring coordinate and find a stationary solution, that is, the black ring solution analytically.

\subsection{Ring coordinate embedding}

The leading order metric has the following asymptotic form at $\sR\gg 1$
%
\begin{eqnarray}
&&
ds^{2}|_{\sR\gg 1} = -dv^{2} +2\left( \frac{H(z)}{\sqrt{1-H'(z)^{2}}}dv-u^{(0)}_{a}(z)\frac{dx^{a}}{n} \right)dr \notag \\
&&~~~~~~~~~~~~~~~~~~~~~~~~~~~~~~
+r^{2}dz^{2}+r^{2}G(z)^{2}d\Phi^{2} +r^{2}H(z)^{2}d\Omega^{2}_{n}.
\label{AsymLO}
\end{eqnarray}
%
We embed this leading order metric into a flat background in the ring coordinate. The $D=n+4$ dimensional 
flat metric in the ring coordinate is \cite{Emparan:2006mm}
\footnote{
In this note we use the ring coordinate by $(r,\theta)$ in \cite{Emparan:2006mm}. Of course the following analysis can be
done also by using $(x,y)$ coordinate in \cite{Emparan:2006mm}.
}
%
\begin{eqnarray}
ds^{2}=-dt^{2} +\frac{R^{2}}{(R+r\cos{\theta})^{2}}\Biggl[
\frac{R^{2}dr^{2}}{R^{2}-r^{2}}
 + (R^{2}-r^{2})d\Phi^{2}
+r^{2}(d\theta^{2} +\sin^{2}{\theta}d\Omega^{2}_{n})
\Biggr],
\label{ringm}
\end{eqnarray}
%
where $0\leq r\leq R$, $0\leq\theta\leq \pi$ and $0\leq\Phi\leq 2\pi$. $R$ is a ring radius. 
$r=0$ is the origin of the ring coordinate. The asymptotic infinity and the axis of $\Phi$-rotation are at $r=R$.     
In this ring coordinate $r=\text{const.}$ surface has a topology of $S^{1}\times S^{n+1}$. $\theta=0$ is an inner equatorial plane, and $\theta=\pi$ is an outer equatorial plane.  
Remembering the definition of $\sR=(r/r_{0})^{n}$, we can embed the leading order induced metric on $\sR=\text{const.}$ surface into the flat background in this ring coordinate 
by $r=r_{0}$ where $r_{0}$ is a constant. As done in the previous section we set to $r_{0}=1$.   
Then, comparing eqs. (\ref{AsymLO}) and (\ref{ringm}) on $r=1$ surface, we find that the embedding gives following identifications
%
\begin{eqnarray}
H(z)=\frac{R  \sin{\theta}}{R+\cos{\theta}},~~
G(z)=\frac{R \sqrt{R^{2}-1}}{R+\cos{\theta}},~~
\frac{d\theta}{dz}=\frac{R+\cos{\theta}}{R }.
\label{BR}
\end{eqnarray}
%
It is easy to confirm that this identification actually satisfies eq. (\ref{HcondLO}).
We can calculate the surface gravity, $\kappa=n\hat{\kappa}$, by these identifications as
%
\begin{eqnarray}
\hat{\kappa}=\frac{\sqrt{1-H'(z)^{2}}}{2H(z)}=\frac{\sqrt{R^{2}-1}}{2R}.
\label{kappa} 
\end{eqnarray}
%
The embedded solution is the black ring solution because the horizon topology is now $S^{1}\times S^{n+1}$. From $R\geq r$, the ring radius should be 
larger than unity $R\geq 1$. $R\simeq 1$ corresponds to the fat black ring, and thin black ring is described by $R\gg 1$. 

Using identifications (\ref{BR}) and the surface gravity (\ref{kappa}), we obtain the leading order effective equations
for the black ring from eqs. (\ref{Deq1}), (\ref{Deq2}) and (\ref{Deq3}) as
%
\begin{eqnarray}
&&
\pv p_{v} +\frac{(R+y)(1+Ry)}{R\sqrt{R^{2}-1}}\partial_{y}p_{v}-\frac{(R+y)^{2}}{R(R^{2}-1)^{3/2}}\pp^{2}p_{v} \notag\\
&&~~~~~~~~~~~~~~~~~~~~~
+\frac{(R+y)^{2}}{R^{2}(R^{2}-1)}\pp p_{\phi} +\frac{1+Ry}{R\sqrt{1-y^{2}}}p_{z}=0,
\label{Deq1BR}
\end{eqnarray}
%
%
\begin{eqnarray}
&&
\pv p_{\phi} +\frac{(R+y)(1+Ry)}{R\sqrt{R^{2}-1}}\partial_{y}p_{\phi}-\frac{(R+y)^{2}}{R(R^{2}-1)^{3/2}}\pp^{2}p_{\phi} \notag\\
&&~~~~~~~~
+\frac{(R+y)^{2}}{R^{2}(R^{2}-1)}\pp\Biggl[\frac{p_{\phi}^{2} }{p_{v}}\Biggr]-\frac{1+2Ry+R^{2}}{R^{2}-1}\pp p_{v} \notag \\
&&~~~~~~~~
+\frac{1+Ry}{R\sqrt{1-y^{2}}}\frac{p_{z}p_{\phi}}{p_{v}}+\frac{2(1+Ry)}{R\sqrt{R^{2}-1}}p_{\phi}=0
\label{Deq2BR}
\end{eqnarray}
%
and
%
\begin{eqnarray}
&&
\pv p_{z} +\frac{(R+y)(1+Ry)}{R\sqrt{R^{2}-1}}\partial_{y}p_{z}-\frac{(R+y)^{2}}{R(R^{2}-1)^{3/2}}\pp^{2}p_{z} \notag\\
&&~~~~~~~~
-\frac{(R+y)\sqrt{1-y^{2}}}{R}\partial_{y} p_{v} +\frac{(R+y)^{2}}{R^{2}(R^{2}-1)}\pp\Biggl[ \frac{p_{z}p_{\phi}}{p_{v}} \Biggr]
+\frac{1+Ry}{R\sqrt{1-y^{2}}}\frac{p_{z}^{2}}{p_{v}} \notag \\
&&~~~~~~~~
-\frac{(R+y)^{2}\sqrt{1-y^{2}}}{R^{3}(R^{2}-1)}\frac{p_{\phi}^{2}}{p_{v}}
+\frac{2(R+y)^{2}\sqrt{1-y^{2}}}{R^{2}(R^{2}-1)^{3/2}}\pp p_{\phi}
+\frac{(R+y)\sqrt{1-y^{2}}}{R^{2}-1}p_{v} \notag \\
&&~~~~~~~~
+\frac{2+2Ry-y^{2}+R^{2}(2y^{2}-1)}{R\sqrt{R^{2}-1}(1-y^{2})}p_{z}=0.
\label{Deq3BR}
\end{eqnarray}
%
Here we introduced a coordinate $y$ defined by
%
\begin{eqnarray}
y=\cos{\theta}.
\end{eqnarray}
%
Our effective equations (\ref{Deq1BR}), (\ref{Deq2BR}) and (\ref{Deq3BR}) 
describe non-linear dynamical deformations of the black ring from thin $R\gg 1$ to not-thin $R>1$ region.

\subsection{Black ring solution}

The black ring solution is obtained as the stationary solution of the effective equations (\ref{Deq1BR}), (\ref{Deq2BR}) and (\ref{Deq3BR}). 
The stationary solution is given by
%
\begin{eqnarray}
p_{v}=e^{P(y)},~~p_{a}=p_{a}(y). 
\end{eqnarray}
%
As done in eq. (\ref{psol}), we find from eqs. (\ref{Deq1BR}) and (\ref{Deq2BR}) 
%
\begin{eqnarray}
p_{z}(y) = -\frac{(R+y)\sqrt{1-y^{2}}}{\sqrt{R^{2}-1}}p_{v}'(y),~~
p_{\phi} = \hat{\Omega}_{H}\frac{R^{2}(R^{2}-1)}{(R+y)^{2}}p_{v}(y). 
\end{eqnarray}
%
Furthermore eq. (\ref{Deq3BR}) gives an equation for $p_{v}=e^{P(y)}$ as eq. (\ref{PeqS})
%
\begin{eqnarray}
P''(y) +\frac{2}{R+y}P'(y) -\frac{R}{(R+y)(1+Ry)}+\hat{\Omega}^{2}_{\tH}\frac{R^{2}(R^{2}-1)^{2}}{(R+y)^{4}(1+Ry)}=0.
\label{PeqBR}
\end{eqnarray}
%
This equation contains a pole at $y=-1/R$ in the source term. The solution can have a singular behavior by the source term at $y=-1/R$, too. 
To obtain a regular solution at $y=-1/R$, $\hat{\Omega}_{H}$ should be
%
\begin{eqnarray}
\hat{\Omega}_{H}=\frac{\sqrt{R^{2}-1}}{R^{2}}.
\label{Ocond}
\end{eqnarray}
%
If $\hat{\Omega}_{\tH}$ does not take this value, the function $P(y)$ has a logarithmic divergence at $y=-1/R$. 
Thus, as seen in five dimensional black ring \cite{Emparan:2001wn} and higher dimensional black ring by the blackfold \cite{Emparan:2007wm,Armas:2014bia}, 
the regularity condition determines the horizon angular velocity of the black ring. 
The solution of eq. (\ref{PeqBR}) under the condition (\ref{Ocond}) is obtained analytically as
%
\begin{eqnarray}
P(y) = p_{0} +\frac{d_{0}}{R+y} +\frac{(1+Ry)(1+Ry+2R(R+y)\log{(R+y)})}{2R^{2}(R+y)^{2}},
\label{Psol}
\end{eqnarray}
%
where $p_{0}$ and $d_{0}$ are integration constants. They are degree of freedom associated with trivial deformations and 
do not affect physical properties of black rings. $p_{0}$ is the $1/n$ redefinition of $r_{0}$, 
and $d_{0}$ comes from the redefinition of $\phi$ coordinate of the ring coordinate as discussed in Appendix \ref{A}.  
Summarizing above results, we found the black ring solution at the leading order of large $D$ expansion by the metric 
%
\begin{eqnarray}
&&
ds^{2} = -\left(1-\frac{p_{v}(z)}{\sR}\right)dv^{2} + 2\left( \frac{dv}{2\hat{\kappa}}-u_{a}^{(0)}(z)\frac{dx^{a}}{n}\right) dr \notag \\
&&~~~~~~~~~~~~~~~~~~~~~~
-2\left( \frac{R+y}{2\hat{\kappa}R}\frac{p_{v}'(z)}{\sR}\frac{dz}{n} 
+\hat{\Omega}_{\tH}\frac{R^{2}(R^{2}-1)}{(R+\cos{\theta)^{2}}}\frac{p_{v}(z)}{\sR}\frac{d\phi}{n} \right)dv \notag \\
&&~~~~
+ r^{2}dz^{2} +\hat{\Omega}_{\tH}\frac{8\hat{\kappa}^{2}R}{R+\cos{\theta}}\frac{p_{v}'(z)}{\sR}\frac{dzd\phi}{n^{2}} \notag \\
&&~~~~
+ r^{2}\frac{R^{2}(R^{2}-1)}{(R+\cos{\theta})^{2}}\left( 1-\frac{2R(R+\cos{\theta})}{R^{2}-1}\frac{\log{\sR}}{n}
+ \hat{\Omega}^{2}_{\tH}\frac{R^{2}(R^{2}-1)}{(R+\cos{\theta})^{2}} \frac{p_{v}(z)}{n\sR} \right)\frac{d\phi^{2}}{n} \notag \\
&&~~~~ 
+ \frac{r^{2}R^{2}\sin^{2}{\theta}}{(R+\cos{\theta})^{2}}d\Omega^{2}_{n},
\label{LOsol}
\end{eqnarray}
%
where $\sR=(r/r_{0})^{n}$ with $r_{0}=1$. The coordinates $z$ and $\theta$ are related by eq. (\ref{BR}). $\hat{\kappa}$ and $\hat{\Omega}_{\tH}$ are
given in eqs. (\ref{kappa}) and (\ref{Ocond}). The function $p_{v}=e^{P(y)}$ is the solution of the equation (\ref{PeqBR}), and it is
obtained in eq. (\ref{Psol}) with $y=\cos{\theta}$. Our black ring solution breaks down at $R=1$, and the very fat black ring
cannot be captured by our large $D$ solution \footnote{
A very fat black ring means the solution with $R=1+O(n^{-1})$. 
}. However, the black ring with not so large ring radius can be described by eq. (\ref{LOsol}). 
So we can study the properties of not only the thin black ring but also not-thin black ring $R>1$ by using our solution (\ref{LOsol}).

\subsection{Quasinormal modes}

We investigate quasinormal modes of the black ring solution. The quasinormal modes are obtained by 
perturbation analysis of the effective equations around the black ring solution. The perturbation ansatz is
%
\begin{eqnarray}
&&
p_{v}(v,y,\phi) = e^{P(y)} \left( 1+ \epsilon e^{-i\omega v}e^{i m\phi}F_{v}(y) \right), \\
&&
p_{z}(v,y,\phi)=-\frac{(R+y)\sqrt{1-y^{2}}}{2\hat{\kappa}R}p_{v}'(y)\left( 1 + \epsilon e^{-i\omega v}e^{i m\phi}F_{z}(y) \right), \\
&&
p_{\phi}(v,y,\phi)= \frac{R(R^{2}-1)^{3/2}}{(R+y)^{2}}p_{v}(y) \left( 1+ \epsilon e^{-i\omega v}e^{i m\phi}F_{\phi}(y) \right),
\end{eqnarray}
%
where we used eq. (\ref{Ocond}). 
There is one remark on the quantum
number $m$ associated with $\partial_{\phi}$. In the ring coordinate (\ref{ringm}) the coordinate $\Phi$ has the range of $0\leq\Phi\leq 2\pi$. 
Thus the quantum number $m_{\Phi}$ associated with $\partial_{\Phi}$ is quantized as $m_{\Phi}=0,\pm1,\pm2,...$. From the relation (\ref{phiRe}),
these quantum numbers are related by 
%
\begin{eqnarray}
m=\frac{m_{\Phi}}{\sqrt{n}}. 
\label{mrel}
\end{eqnarray}
%
So $m$ in the perturbations can take non-integer values in general. 

Perturbing the effective equations (\ref{Deq1BR}), (\ref{Deq2BR}) and (\ref{Deq3BR}) with respect to $\epsilon$, 
we obtain perturbation equations for $F_{v}(y)$, 
$F_{z}(y)$ and $F_{\phi}(y)$. The perturbation equations have a pole at $y=-1/R$ again. 
To solve the perturbation equations we should impose regularity conditions. If we specify the behavior of the perturbation fields at the pole
$y=-1/R$ as the regularity condition by
%
\begin{eqnarray}
F_{v}(y) \propto (1+Ry)^{\ell}\left(
1+O(1+R y)
\right),
\label{polecond}
\end{eqnarray}
%
where $\ell$ is a non-negative integer, we get one non-trivial condition for the frequency $\omega$ as
%
\begin{eqnarray}
&&
\frac{1}{\sqrt{R^{2}-1}(m^{2}+imR+\ell R^{2})-iR^{3}\omega}\Bigl[
R^{9}\omega^{3} + iR^{6}\sqrt{R^{2}-1}\Bigl( 3m^{2}+3imR \notag \\
&&~~~
+(3\ell-2)R^{2} \Bigr)\omega^{2}
-R^{3}(R^{2}-1)\Bigl( 3m^{4}+6im^{3}R+2(3\ell-4) m^{2}R^{2} \notag \\
&&~~~+2i(3\ell-2)m R^{3}+3(\ell-1)\ell R^{4}\Bigr)\omega 
-i(R^{2}-1)^{3/2}\Bigl( m^{6}+3im^{5}R +3(\ell-2)m^{4}R^{2} \notag \\
&&~~~
+6i(\ell-1)m^{3}R^{3} 
-(4-7\ell+3\ell^{2})m^{2}R^{4}+3i\ell(\ell-1)mR^{5} +\ell^{2}(\ell-1)R^{6} \Bigr)
\Bigr] \notag \\
&&~~~
=0.
\label{QNMcondBR}
\end{eqnarray}
%
This is the quasinormal mode condition for the black ring. One may feel strange about this derivation since we derive the quasinormal mode condition
from the local condition (\ref{polecond}) at $y=-1/R$. If the condition (\ref{QNMcondBR}) is satisfied, the perturbation actually becomes regular at $y=-1/R$, 
but its global structure, {\it e.g.}, of $F_{v}(y)$, is still unknown. Furthermore the quantum number $\ell$ associated with the harmonics in the 
ring coordinate should be given by the global 
solution of $F_{v}(y)$. But the quantum number is introduced by the local condition (\ref{polecond}) in our derivation at the large $D$ limit.  
This localization of the quantum number associated with the harmonics at the large $D$ limit has been observed in the spherical harmonics and spheroidal harmonics in \cite{Suzuki:2015iha}. 
So we expect that the same feature would appear in the harmonics in the ring coordinate. It is interesting to investigate this property in detail, 
although we do not pursue this structure in this paper. In the following we give results derived from eq. (\ref{QNMcondBR}) for non-axisymmetric 
and axisymmetric modes separately.    
 
\paragraph{Non-axisymmetric modes $(m\neq 0)$}
 
The quasinormal mode condition (\ref{QNMcondBR}) can be solved in a simple form for $\ell=0$ by
%
\begin{eqnarray}
\omega^{(\ell=0)}_{\pm}= \frac{\sqrt{R^{2}-1}}{R}\Bigl[ \hat{m} \pm i\hat{m}(1 \mp \hat{m}) \Bigr],~~
\omega^{(\ell=0)}_{0} = \frac{\sqrt{R^{2}-1}}{R}\Bigl[ \hat{m} -i(\hat{m}^{2}-2) \Bigr], 
\label{QNMell0BR}
\end{eqnarray}
%
%
\begin{figure}[t]
 \begin{center}
  \includegraphics[width=65mm,angle=0]{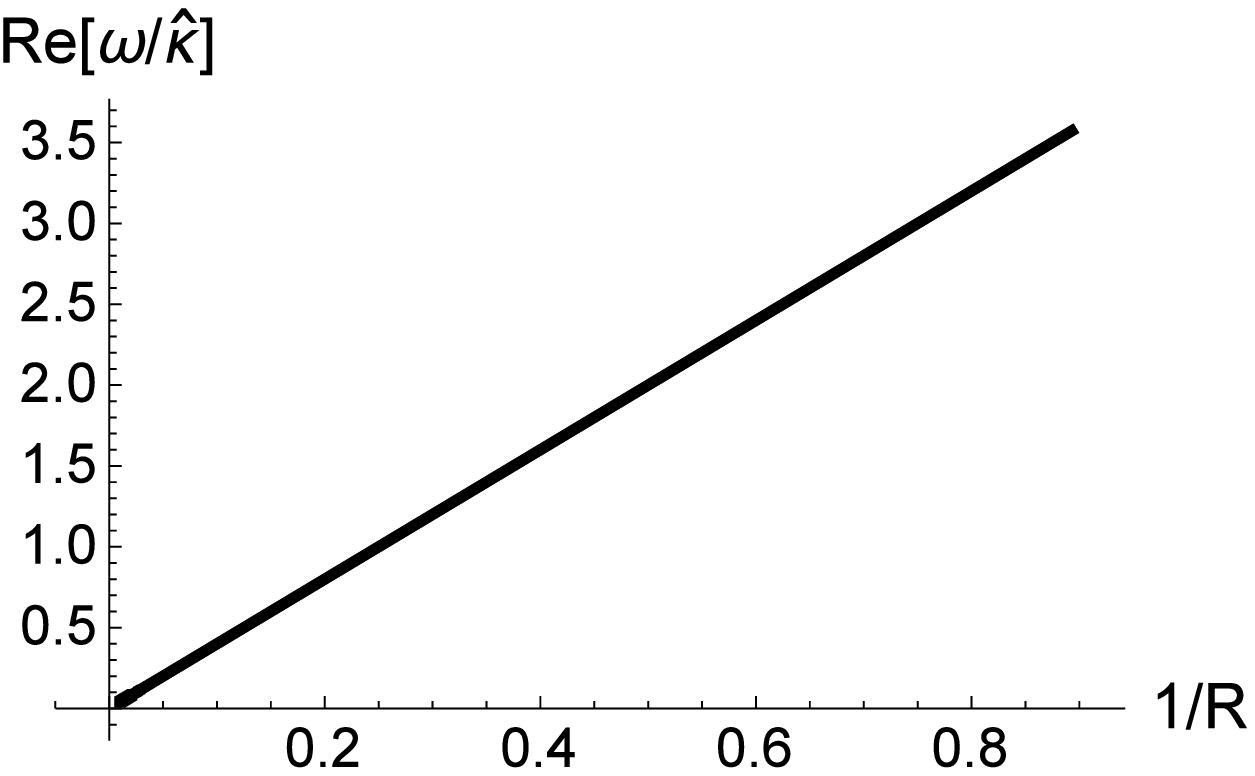}
  \hspace{5mm}
  \includegraphics[width=65mm,angle=0]{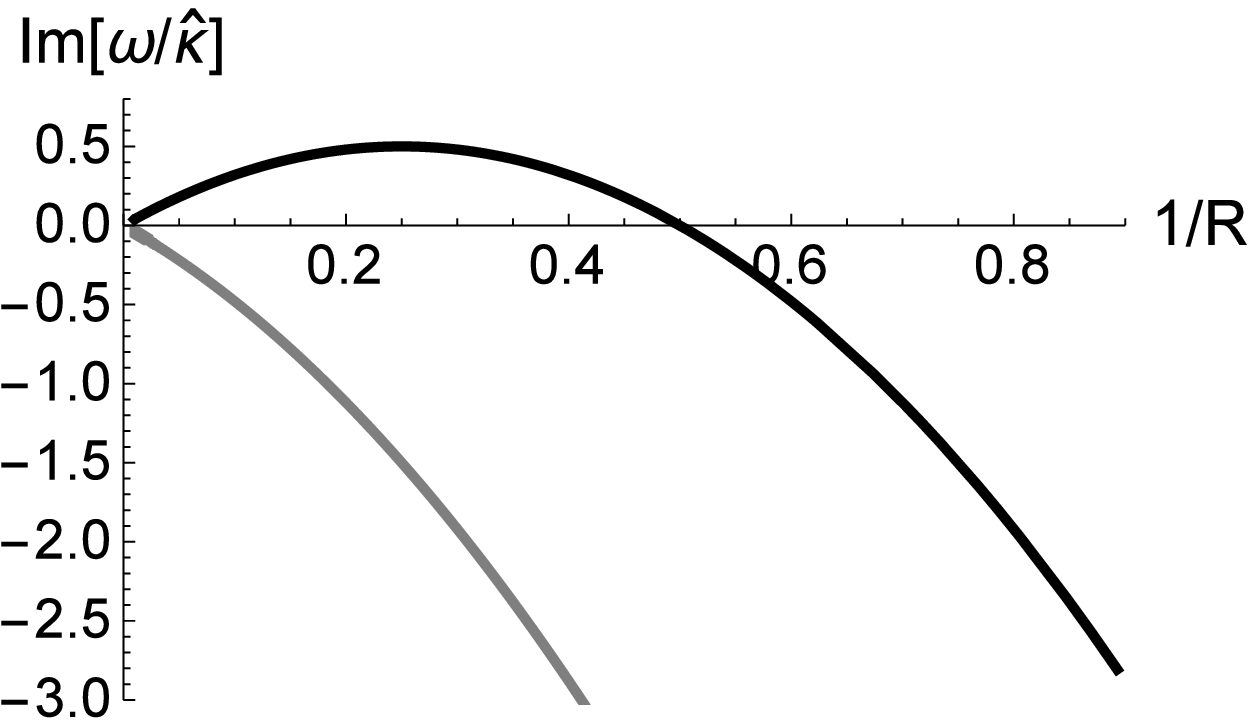}
 \end{center}
 \vspace{-5mm}
 \caption{ Plots of the leading order quasinormal modes with $\ell=0$ and $m=2$, $\omega^{(\ell=0)}_{\pm}$, of the black ring. 
The left panel shows the real part of the frequency normalized by the reduced surface gravity $\hat{\kappa}$. The right panel shows the imaginary part.
The black and gray line represent $\omega^{(\ell=0)}_{+}$ and $\omega^{(\ell=0)}_{-}$ respectively. The real part of $\omega^{(\ell=0)}_{+}$ and 
$\omega^{(\ell=0)}_{-}$ are same. 
$\omega^{(\ell=0)}_{+}$ is unstable for $R>2$.  }
 \label{QNMBR}
\end{figure}
%
where $\hat{m}=m/R$. $\omega^{(\ell=0)}_{+}$ is an instability mode when $R>m$.
$\omega_{\pm}^{(\ell=0)}$ can be understood as the quasinormal mode of the boosted black string as we will see below. 
We regard $\omega_{0}^{(\ell=0)}$ in eq. (\ref{QNMell0BR}) as a gauge mode in this paper
\footnote{
We have not still studied residual gauge of $(v,z,\phi)$ coordinate in detail. The gauge of $r$-coordinate was fixed in eq. (\ref{mansatz}). So we should check the gauge transformation of perturbations to 
check which modes are physical. Actually there are residual gauge. In Appendix \ref{A} we study stationary one of them. It is a bit involved task to eliminate whole gauge modes. So instead we use another simple method to identify physical modes here.  
}. At large radius limit $R\gg 1$ the quasinormal modes
should reproduce the quasinormal modes of the boosted black string. Although $\omega^{(\ell=0)}_{\pm}$ has corresponding modes, $\omega^{(\ell=0)}_{0}$
does not have. So it might be natural to consider that $\omega^{(\ell=0)}_{0}$ is a gauge mode. 
In Figure \ref{QNMBR} we show plots of the quasinormal mode $\omega^{(\ell=0)}_{\pm}$
with $m=2$. $\omega^{(\ell=0)}_{+}$ becomes stable in relatively fat region $R<m$. This plot reproduces the behavior of the numerical
results in \cite{Santos:2015iua}
\footnote{
The perturbation considered in \cite{Santos:2015iua} is the mode with $m_{\Phi}=2$, not $m=2$ as seen in eq. (\ref{mrel}). Thus our result for $m=2$
does not exactly correspond to the mode in \cite{Santos:2015iua}. However, in $D=5 ~(n=1)$, we can expect that $m_{\Phi}=2$ and $m=2$
are not different so much. 
}.  
We could not find any instabilities for $\ell\neq 0$ and $m\neq 0$ mode perturbations. 

Note that the instability mode $\omega^{(\ell=0)}_{+}$ always saturates the superradiant condition
\footnote{
The quantum numbers $m$ and $m_{\Phi}$ has the relation (\ref{mrel}). Then the superradiant factor
is 
%
\begin{eqnarray}
\text{Re}\Bigl[
\omega - m_{\Phi}\Omega_{\tH}
\Bigr]
=
\text{Re}\Bigl[
\omega - m\hat{\Omega}_{\tH}
\Bigr]. \notag
\end{eqnarray}
%
}
%
\begin{eqnarray}
\text{Re}\Bigl[
\omega^{(\ell=0)}_{+} -m\hat{\Omega}_{\tH}
\Bigr]=0.
\end{eqnarray}
%
The nature of this saturation at the large $D$ limit is unclear. The dynamically unstable mode of the Myers-Perry black hole also shows 
the peculiar relation with the superradiant condition at the large $D$ limit \cite{Suzuki:2015iha, Emparan:2014jca}. For equally spin 
Myers-Perry black hole, the coincidence of the onset of the superradiant and dynamically unstable regime was confirmed up to $1/D$ corrections 
\cite{Emparan:2014jca}, 
and it is consistent with the numerical result \cite{Hartnett:2013fba}. For singly rotating Myers-Perry black hole the coincidence was 
also found at the large $D$ limit \cite{Suzuki:2015iha}. However it is known numerically that
their onsets are at different rotation parameters for the singly rotating Myers-Perry black hole in finite dimensions \cite{Shibata:2010wz,Dias:2014eua}. 
Thus the coincidence of the onsets of superradiant and dynamically unstable regime is only the feature at the large $D$ limit for the singly rotating Myers-Perry black hole.  
As we will see below, the saturation of the superradiant condition of the black ring does not always hold at higher order in $1/n$ expansion, 
and we find that onsets of the superradiant and 
dynamically unstable regime are same also for the black ring at the large $D$ limit.

\subparagraph{Endpoints of instability?}

We found non-axisymmetric instabilities of the black ring solution. 
One may ask what its endpoint is. In \cite{Elvang:2006dd} they 
discussed that the instability of the five dimensional thin black ring leads to the fragmentation into black holes by following two reasons. One is that 
the dynamical timescale to release the inhomogeneity by gravitational wave radiations is much longer than the timescale of the black ring instability. 
Thus the inhomogeneity by the instability of the thin black ring will grow in time. 
Another reason is that the fragmenting solution can be more entropic than the black ring. This can be understood by the fact 
that the non-uniform black string is less entropic than localized black holes in five dimensions. Then we expect that going to the fragmenting solution 
is preferable as the endpoint of the instability than going back to the (fatter) stable black ring. So the growth of the inhomogeneity of the black ring
would not stop, and the fragmentation occurs.   
In our case this second reason of the discussion cannot hold. 
The non-uniform black string can be the endpoint of the Gregory-Laflamme instability as seen in \cite{Emparan:2015gva} since the non-uniform black string
becomes more entropic solution in higher dimensions than the critical dimension \cite{Sorkin:2004qq}. Then, also for the black ring instability, 
we can say same thing. In enough higher dimensions the non-uniform solution is more entropic than the fragmenting solutions. So we expect that
the black ring would not go to the fragmentation by the instability. On the other hand the first reason can be applied to our case. Actually the 
inhomogeneity by the black ring instability and its rotation give time-dependent quadrupole moments to gravitational fields. Then the timescale
of gravitational wave radiations by
time-dependent quadrupole moments, $t_{\text{GW}}$, is estimated as done in \cite{Elvang:2006dd} by
%
\begin{eqnarray}
t_{\text{GW}} \sim \frac{1}{\Omega_{\tH}}\frac{R^{n+1}}{\G \mathcal{M}} \sim \frac{R^{n+2}}{\G \mathcal{M}} \sim e^{n/2}R\left(\frac{R}{r_{0}}\right)^{n},
\end{eqnarray}
%
where the exponential factor $e^{n/2}$ comes from $\Omega_{n}$ in $\G\mathcal{M}$ as seen in eq. (\ref{MLO}). 
We restored the black ring thickness $r_{0}$ for usefulness.
The timescale of the black ring instability, $t_{\text{BR}}$, is estimated by the quasinormal mode frequency as
%
\begin{eqnarray}
t_{\text{BR}}= \left(\text{Im}\left[\omega^{(\ell=0)}_{+} \right] \right)^{-1}\sim r_{0}.
\end{eqnarray}
%
So $t_{\text{BR}}$ is $O(1)$ quantity in $1/n$ expansion. Then, remembering that our large $D$ black ring solution satisfies $R>r_{0}$, 
the ratio of timescales is exponentially small in $n$ at large $D$
%
\begin{eqnarray}
\frac{t_{\text{BR}}}{t_\text{GW}} \sim e^{-n/2}\left(\frac{R}{r_{0}}\right)^{-n-1} \ll O(1).
\end{eqnarray}
%
This implies that inhomogeneity by the black ring instability cannot be dissipated by the gravitational wave
radiation to the infinity in $1/n$ expansion. But the inhomogeneous black ring solution would not fragment into small
black holes unlike in five dimensions as discussed above. Thus the inhomogeneity would stop at some point like the non-uniform black string 
observed in \cite{Emparan:2015gva}. Then we reach the conclusion for the endpoint of the black ring instability
that the black ring evolves to a non-uniform black ring, NUBR, as a metastable solution at large $D$
\footnote{
The author thanks Roberto Emparan for the suggestion of this possibility. 
}. The metastable solution at large $D$
means that the solution is stationary in $1/n$ expansion, but it is not stationary in $O(e^{-n/2})$ by the gravitational 
wave emissions. The stationary solution with inhomogeneities along rotating direction is prohibited by the rigidity theorem
\cite{Hawking:1971vc,Hollands:2006rj}. Our statement for the endpoint of the black ring instability is consistent with the rigidity theorem
since NUBR is not stationary in $O(e^{-n/2})$. Such exponentially suppressed evolutions cannot be described by our effective equations
(\ref{Deq1BR}), (\ref{Deq2BR}) and (\ref{Deq3BR}) by $1/n$ expansions. It is interesting to investigate this possibility by solving the 
effective equations for the black ring (\ref{Deq1BR}), (\ref{Deq2BR}) and (\ref{Deq3BR}) directly as the investigation of the endpoint 
of the Gregory-Laflamme instability in \cite{Emparan:2015gva}. If there is a metastable solution such as NUBR, 
we can find a stationary and non-axisymmetric solution of the effective equations. 

This statement for the endpoint of the black ring instability is only for the leading order result in $1/n$ expansion. 
If we include $1/n$ corrections to the effective equations, the results show the dimensional dependences, and 
we can observe the critical dimension in which the stability of NUBR would change. Actually we have investigated the 
critical dimension of the non-uniform black string by considering $1/n$ corrections \cite{Suzuki:2015axa}. So our
conjecture that the endpoint of the black ring instability is a stable NUBR is valid only at the leading order in $1/n$
expansion. 

\paragraph{Axisymmetric modes $(m = 0)$}

For $m=0$ modes, we can also solve eq. (\ref{QNMcondBR}) in a simple form as
%
\begin{eqnarray}
\omega^{(m=0)}_{\pm}=\pm{\sqrt{\ell-1}}-i(\ell-1). 
\label{QNMm0BR}
\end{eqnarray}
%
Note that the numerator of eq. (\ref{QNMcondBR}) is a cubic algebraic equation of $\omega$, but one solution of the numerator equation is canceled with the denominator of eq. (\ref{QNMcondBR}) 
for $m=0$ modes. So there are only two modes as eq. (\ref{QNMm0BR}). The quasinormal mode frequencies  (\ref{QNMm0BR}) are same with one of the
Schwarzschild black hole at large $D$ \cite{Emparan:2014aba}. This is because we consider the decoupled mode excitations. In the decoupled 
mode excitation, the dynamics of perturbations is determined almost locally on the horizon. For $m=0$ modes, the perturbation 
does not have interactions along $\phi$ direction. Thus the perturbation feels the horizon as one of the $D-1$ dimensional Schwarzschild black hole, 
and its quasinormal mode becomes one of the Schwarzschild black hole. 

The axisymmetric perturbation does not have any instabilities at the leading order of $1/n$ expansion. 
Thus the instability found in \cite{Elvang:2006dd,Figueras:2011he}
for the fat black ring against axisymmetric perturbations is not contained in our setup. Such instability occurs due to interactions
between horizons of the black ring, so it may be non-decoupled mode instability. However, by including $1/n$ corrections, 
we find a suggestion for the axisymmetric instabilities of the black ring as we will see later. 

\subsection{Black string limit}

As one useful observation let us see the large ring radius limit $R \gg 1$ of the black ring solution 
found above. This limit gives the expression of the black ring as a boosted black string \cite{Emparan:2007wm}. 
At the large radius limit we can set the behavior of $P(y)$ in eq. (\ref{Psol}) to
%
\begin{eqnarray}
P(y) = O(1/R),
\end{eqnarray}
%
by $p_{0}=0$ and $d_{0}=0$. 
Then we get the large radius limit of the black ring solution as
%
\begin{eqnarray}
ds^{2} = -\left( 1-\frac{1}{\sR} \right) dv^{2} +2dvdr -\frac{1}{\sqrt{n}\sR}dv dx 
+\left( 1-\frac{1}{n\sR} \right)dx^{2}+ r^{2}d\Omega^{2}_{n}, 
\label{boosttrans}
\end{eqnarray}
%
where we defined the black string direction $dx$ by $dx=Rd\phi/\sqrt{n}=Rd\Phi$. This large radius limit solution is
actually regarded as the large $D$ limit metric of the boosted string with the boost velocity 
%
\begin{eqnarray}
\sinh{\alpha} = \frac{1}{\sqrt{n}}, 
\label{boost}
\end{eqnarray}
%
as found in \cite{Emparan:2007wm}. Using this boost relation we can reproduce the large radius limit 
of the quasinormal modes, $\omega^{(\ell=0)}_{\pm}$, of the black ring from the quasinormal modes of the black string. 
The quasinormal modes of the black string, $\omega^{\text{BS}}_{\pm}$ was obtained in \cite{Emparan:2013moa} 
by large $D$ expansion. For the perturbation $\sim e^{-i\omega v}e^{ikx}$, the leading order result is
%
\begin{eqnarray}
\omega^{\text{BS}}_{\pm}=\pm i \hat{k}(1\mp\hat{k}),
\end{eqnarray}
%
where $\hat{k}=k/\sqrt{n}$. Now the boost transformation on the black string is acting by
%
\begin{eqnarray}
dv \rightarrow \cosh{\alpha}dv -\sinh{\alpha}dx,~~
dx \rightarrow \cosh{\alpha}dx -\sinh{\alpha}dv.
\end{eqnarray}
%
This transformation on $\omega^{\text{BS}}_{\pm}$ gives quasinormal modes of the boosted black string, 
$\omega^{\text{bBS}}_{\pm}$, as
%
\begin{eqnarray}
\omega^{\text{BS}}_{\pm} \rightarrow \omega^{\text{bBS}}_{\pm}=k\sinh{\alpha}+\omega^{\text{BS}}_{\pm}\cosh{\alpha}.
\end{eqnarray}
%
Using eq. (\ref{boost}) and $\hat{k}=k/\sqrt{n}$, we obtain the quasinormal modes of the boosted black string 
with the boost velocity (\ref{boost}) as

%
\begin{eqnarray}
\omega^{\text{bBS}}_{\pm}=\hat{k}\pm i\hat{k}(1\mp\hat{k}). 
\end{eqnarray}
%
This quasinormal modes precisely reproduce the large radius limit of quasinormal modes of the black ring
by identifying $\hat{k}=\hat{m}$
\footnote{
The boost transformation is one by the exact symmetry of the black string. Thus we can obtain the quasinormal mode
of the boosted black string from the black string. As for the singly rotting Myers-Perry black hole, the boost transformation
comes from an approximation symmetry appearing only at the leading order of large $D$ limit. Thus quasinormal modes
of the Myers-Perry black hole do not follow boost transformation rules.  
}.  
In Appendix \ref{B} we give the direct derivation of the quasinormal modes of the boosted black string from the effective equations.

\subsection{$1/D$ corrections}

We can obtain $1/n$ corrections to results obtained above by solving the effective equations up to $O(1/n)$.
In the following we set to $p_{0}=0$ and $d_{0}=0$ for $P(y)$ in eq. (\ref{Psol}).
$p_{v}(z)$ and $p_{\phi}(z)$ for the stationary solution
are
%
\begin{eqnarray}
p_{v}(y)=e^{P(y)}\Biggl[ 1+\frac{P^{(1)}(y)}{n} \Biggr],
\end{eqnarray}
%
%
\begin{eqnarray}
&&
p_{\phi}(y)=\frac{R(R^{2}-1)^{3/2}}{(R+y)^{2}}e^{P(y)} \notag \\
&&~~~\times
\Biggl[ 1+
\frac{1}{n} \Bigl(
P^{(1)}(y) -\frac{2(1+Ry)}{R^{2}-1}P(y)
+\frac{\log{(R-R^{-1})}}{R^{2}}
\Bigr)
\Biggr],
\end{eqnarray}
%
where $P^{(1)}(y)$ is given by
%
\begin{eqnarray}
&&
P^{(1)}(y) = p_{1}+\frac{d_{1}}{R+y} +\frac{(R^{2}-1)^{2}}{R^{4}(R+y)^{2}}\log{(R-R^{-1})}\notag \\
&&~~~~~
-\frac{2(1+Ry)}{R^{3}(R+y)}\text{Li}_{2}\left( \frac{1+Ry}{R^{2}-1} \right) 
+\frac{(R^{2}-1)(y-R(1-2y^{2}))}{2R^{3}(R+y)^{2}}(\log{(R+y)})^{2}\notag \\
&&~~~~~
-\frac{1}{4R^{4}(R^{2}-1)(R+y)^{4}}\Bigl[
3+12Ry+2y^{2}+2R^{8}(4-3y^{2})+2R^{7}y(5-4y^{2}) \notag \\
&&~~~~~~~~~
-R^{2}(5-6y^{2})-12R^{5}y(1+y^{2})
-R^{6}(12-8y^{2}+5y^{4}) 
-2R^{3}y(10+y^{4})\notag \\
&&~~~~~~~~~~
+R^{4}(4-30y^{2}-5y^{4})
\Bigr] \notag \\
&&~~~~~
+\frac{\log{(R+y)}}{R^{3}(R^{2}-1)(R+y)^{3}}\Bigl[
-2+R^{7}y-y^{2}+3R^{2}(1-2y^{2})+R^{6}(2+y^{2})\notag \\
&&~~~~~~~~~~
+3R^{3}y(2+y^{2})-2Ry(3+y^{2}) 
+3R^{5}y(1+y^{2}) 
-R^{4}(2-12y^{2}+y^{4})\notag \\
&&~~~~~~~~~~
+2(1+Ry)(R^{2}-1)(R+y)^{2}\log{(R-R^{-1})}
\Bigr].
\end{eqnarray}
%
$p_{1}$ and $d_{1}$ are integration constants of $1/n$ corrections. $\text{Li}_{2}(x)$ is the polylogarithm function. 
This solution has the surface gravity
%
\begin{eqnarray}
\kappa = n\frac{\sqrt{R^{2}-1}}{2R}\Biggl[
 1 - \frac{1}{2n}+O(n^{-2})
\Biggr], 
\end{eqnarray}
%
and the horizon angular velocity 
%
\begin{eqnarray}
\Omega_{H}=\frac{1}{\sqrt{n}}\frac{\sqrt{R^{2}-1}}{R^{2}} \Biggl[
1-\frac{1}{n}\frac{R^{2}-2\log{(R-R^{-1})}}{2R^{2}}
+O(n^{-2})\Biggr].
\end{eqnarray}
%
The horizon angular velocity up to $O(1/n)$ can be obtained by imposing the regularity of $p_{v}(z)$
at the pole $y=-1/R$ up to $O(1/n)$. We can see that this horizon angular velocity reproduces eq. (\ref{BFformula}) by the blackfold method
at the large radius limit up to $O(1/n)$. 
The surface gravity has the following large radius limit
%
\begin{eqnarray}
\lim_{R\rightarrow \infty}\kappa &=& \frac{n}{2}\Bigl[
1-\frac{1}{2n} +O(n^{-2})
\Bigr] \notag \\
&=&
\frac{n}{2\cosh{\alpha}}\left( 1+O(n^{-2}) \right),
\end{eqnarray}
%
Then this expression implies that the surface gravity is reproduced by the boosted black string with the 
boost velocity (\ref{boost}) up to $1/n$ corrections. We can also compare the $1/n$ corrections 
of our results with the $1/R^{2}$ corrections by the blackfold method \cite{Armas:2014bia} to the
horizon angular velocity. However, in this comparison, there are some subtle things such as a definition of $R$, 
so the comparison is not clear. The fact that $1/n$ corrections do not contain $1/R$ contribution to $\Omega_{H}$
is consistent with the result in \cite{Emparan:2007wm}. 

Finally we give $1/n$ corrections to quasinormal modes of the black ring. 
$\omega^{(\ell=0)}_{\pm}$ up to $1/n$
corrections is
%
\begin{eqnarray}
\omega_{\pm}^{(\ell=0)} = \frac{\sqrt{R^{2}-1}}{R}\Biggl[
\hat{m}\pm i\hat{m}(1\mp\hat{m})+\frac{\delta\hat{\omega}^{(\ell=0)}_{\pm}}{n}
\Biggr],
\end{eqnarray}
%
where
%
\begin{eqnarray}
&&
\delta\hat{\omega}^{(\ell=0)}_{\pm}=\frac{1}{2\hat{m}R^{2}}\Bigl[
\hat{m}^{2}(R^{2}-4+4R^{2}\hat{m}^{2})+i\hat{m}(2+(R^{2}+16)\hat{m}^{2}+8\hat{m}^{4}) \notag \\
&&~~~~~~~~~~~~~~~
+2\hat{m}^{2}(1-2i\hat{m})\log{(R-R^{-1})}
\mp \Bigl(
2\hat{m}(2+(3R^{2}-4)\hat{m}^{2}) \notag \\
&&~~~~~~~~~~~~~~~
+i(2+(3R^{2}+1)\hat{m}^{2}-2(R^{2}-10)\hat{m}^{4}-2\hat{m}^{2}\log{(R-R^{-1})})
\Bigr)
\Bigr].
\label{QNMBRNLOell0}
\end{eqnarray}
%
At the large radius limit $R\rightarrow \infty$ with fixed $\hat{m}$, this quasinormal mode reduces to one of the boosted black string 
with the boost parameter given in eq. (\ref{boost}) as seen in Appendix \ref{B}. Thus we could confirm that the blackfold analysis by \cite{Emparan:2007wm}
is correct up to $1/n$ corrections. 

From this $1/n$ correction of $\omega^{(\ell=0)}_{+}$, we find that the black ring becomes unstable against non-axisymmetric perturbations 
for $R>R_{D}$ where
%
\begin{eqnarray}
R_{D} = m\Biggl[
1 -\frac{1}{n}\frac{3 - 2\log{(m-m^{-1})} }{2m^{2}}  
+O(n^{-2})\Biggr]. 
\label{Rthr}
\end{eqnarray}
%
%
\begin{figure}[t]
 \begin{center}
  \includegraphics[width=65mm,angle=0]{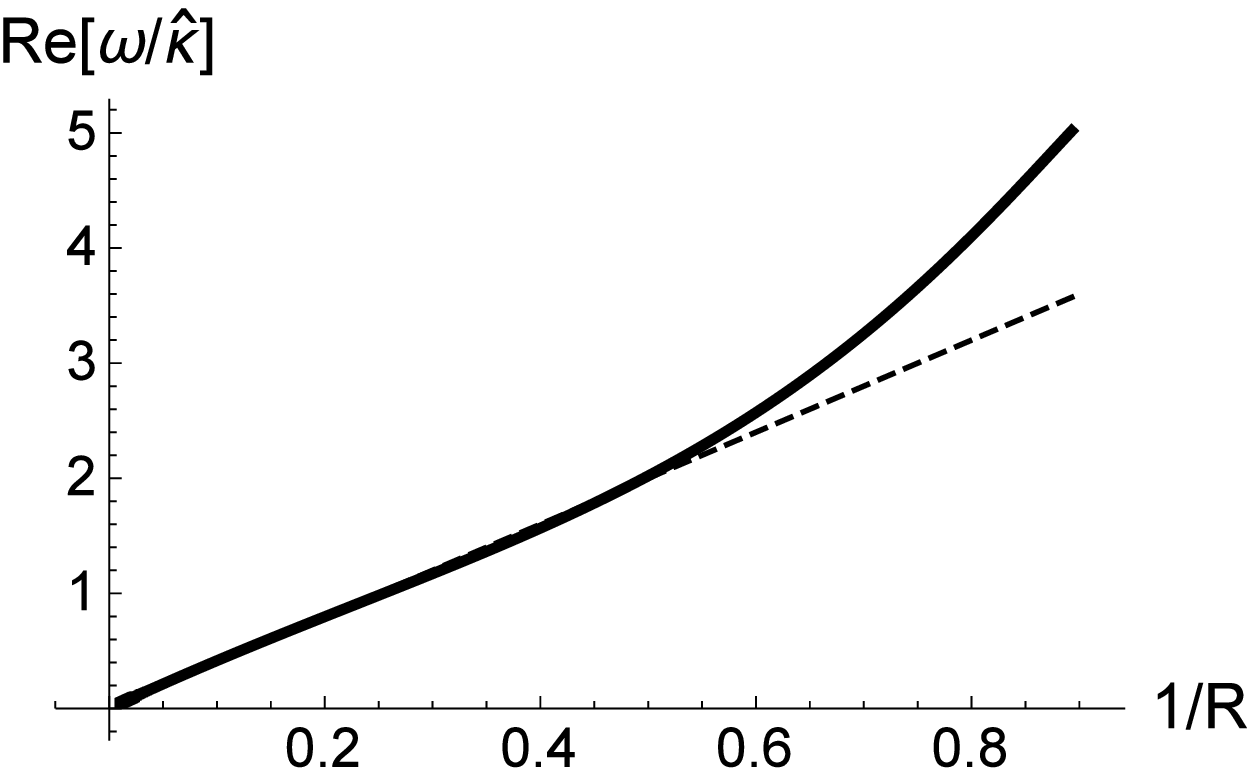}
  \hspace{5mm}
  \includegraphics[width=65mm,angle=0]{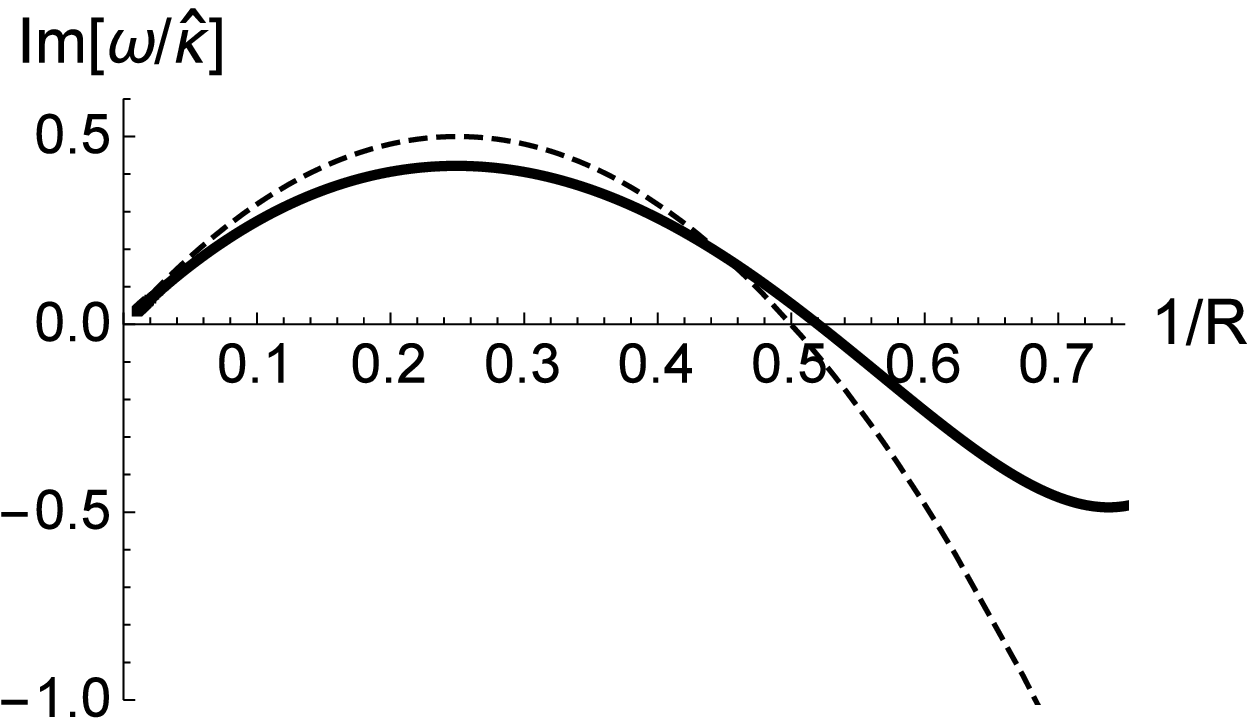}
 \end{center}
 \vspace{-5mm}
 \caption{ Plots of the quasinormal mode $\omega^{(\ell=0)}_{+}$ with $\ell=0$ and $m=2$ in $n=10$ of the black ring up to $O(1/n)$ corrections. 
The left panel shows the real part of the frequency normalized by the reduced surface gravity. The right panel shows the imaginary part.
The dashed line represents the leading order result. The thick line is the result with $O(1/n)$ corrections.
}
 \label{QNMBRNLO}
\end{figure}
%
At the large radius limit this threshold ring radius $R_{D}$ gives the marginally stable "wave number $\hat{m}_{D}$" as
%
\begin{eqnarray}
\hat{m}_{D}\equiv \frac{m}{R_{D}}=1+O(n^{-2}, R^{-2}). 
\label{mD}
\end{eqnarray}
%
This corresponds to the Gregory-Laflamme mode of the boosted black string with the boost velocity (\ref{boost}) as seen in Appendix \ref{B}. 
In Figure \ref{QNMBRNLO} we show the plots of the quasinormal mode frequency $\omega^{(\ell=0)}_{+}$ with $m=2$ in $n=10$ up to the $1/n$ corrections. 
The plots in Figure \ref{QNMBRNLO} show that the results become much closer to numerical results \cite{Santos:2015iua} by $1/n$ corrections.   
At $R=1$ our quasinormal mode formula breaks down due to the term $\log{(R-R^{-1})}$. Thus the quasinormal mode frequency behavior around $R=1$
is not reliable. 

One interesting observation on $1/n$ corrections to quasinormal modes is the relation of the superradiant and dynamically unstable regime. At the leading order the real part of 
the quasinormal mode frequency $\omega^{(\ell=0)}_{+}$ is marginal in the superradiant condition 
%
\begin{eqnarray}
\text{Re}\left[\omega_{+}^{(\ell=0)}-m\hat{\Omega}_{\tH}\right] = O(n^{-1}).  
\end{eqnarray}
%
By including the $1/n$ corrections, we see that the real part of the quasinormal mode frequency $\omega_{+}^{(\ell=0)}$ deviates from the 
superradiant condition. Actually the real part of the frequency does not always satisfy the superradiant condition
%
\begin{eqnarray}
\text{Re}\left[\omega^{(\ell=0)}_{+}-m\hat{\Omega}_{\tH} \right] 
= \frac{2\hat{\kappa}}{n}\frac{(\hat{m}-1)(2-(R^{2}-4)\hat{m}+2R^{2}\hat{m}^{2})}{R^{2}}+O(n^{-2}).
\label{superR}
\end{eqnarray}
%
%
\begin{figure}[t]
 \begin{center}
  \includegraphics[width=80mm,angle=0]{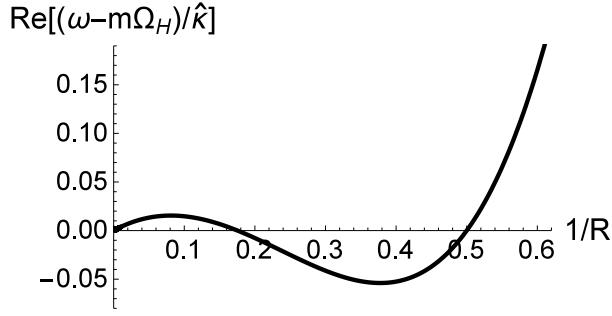}
 \end{center}
 \vspace{-5mm}
 \caption{ The plot of the superradiant factor is shown for the dynamically unstable mode $\omega^{(\ell=0)}_{+}$ with 
$m=2$ and $n=10$. The superradiant factor can be negative and positive in dynamically unstable regime $R>m$, while
it is always positive in stable regime.  
}
 \label{QNMomega}
\end{figure}
%
In Figure \ref{QNMomega} we show the plot of eq. (\ref{superR}) for $n=10$ and $m=2$. The superradiant regime can be seen 
in the dynamically unstable regime $R>m$. But there is also the dynamically unstable mode which 
does not satisfy the superradiant condition at $R>R_{S}>m$,
%
\begin{eqnarray}
R_{S}=\frac{m^{2}+1+\sqrt{1+6m^{2}+m^{4}}}{m}. 
\label{Rs}
\end{eqnarray}
%
Thus the black ring has dynamically unstable modes which satisfy the superradiant condition for $R_{S}>R>m$, and 
the superradiant condition becomes not to be satisfied by unstable modes at  $R>R_{S}$. 
This behavior of the dynamically unstable modes is the striking property of the black ring also seen in numerical results \cite{Santos:2015iua}.
In fact the all known dynamically unstable modes of black holes such as one of the Myers-Perry black holes always satisfy the superradiant condition. 
The physics behind 
this feature is not still clear.  

We can also obtain $1/n$ corrections to the quasinormal modes with $m=0$ in a simple form. The result is
%
\begin{eqnarray}
\omega^{(m=0)}_{\pm}=\frac{\sqrt{R^{2}-1}}{R}\Biggl[
\pm\sqrt{\ell-1}-i(\ell-1) +\frac{\delta\hat{\omega}^{(m=0)}_{\pm}}{n}
\Biggr],
\end{eqnarray}
%
where
%
\begin{eqnarray}
&&
\delta\hat{\omega}^{(m=0)}_{\pm} = \pm \frac{R^{2}\sqrt{\ell-1}}{2(R^{2}-1)} \Biggl[ 
3\ell-5 -\frac{7\ell^{2}-\ell-10}{2R^{2}(\ell-1)}+\frac{2(4\ell-5)}{R^{4}(\ell-1)}
\Biggr] \notag \\
&&~~~~~~~~~~~~~~~~
-i\frac{R^{2}(\ell-1)}{2(R^{2}-1)}\Biggl[
2\ell-5 -\frac{4\ell^{2}-7\ell-10}{2R^{2}(\ell-1)} +\frac{2(2\ell-5)}{R^{4}(\ell-1)}
\Biggr].
\label{QNMNLOm0}
\end{eqnarray}
%
At the large radius limit $R\rightarrow\infty$ the quasinormal mode frequency $\omega^{(m=0)}_{\pm}$ deviates from 
one of Schwarzschild black hole \cite{Emparan:2014aba} in $1/n$ corrections. This deviation can be understood as 
a boost effect of the black string as discussed in Appendix \ref{B}. Around $R=1$  the imaginary part of the $1/n$ 
correction $\delta\hat{\omega}^{(m=0)}_{\pm}$ becomes positive. This might mean that the axisymmetric perturbation becomes
also unstable around $R=1$. Actually we find that there is a marginally stable mode at $R=R_{f}$ as 
%
\begin{eqnarray}
\omega^{(m=0)}_{\pm}\Bigl|_{R=R_{f}}=-\ell\sqrt{\ell-1}+O(n^{-1}),
\end{eqnarray}
%
where $R_{f}$ is the marginal radius for the apparent axisymmetric instability of the black ring given by
%
\begin{eqnarray}
R_{f} = 1 -\frac{1}{n}\frac{\ell}{8(\ell-1)}+O(n^{-2}).
\end{eqnarray}
%
The imaginary part becomes positive at $1>R>R_{f}$, and the axisymmetric perturbation may be unstable there. 
As mentioned above our formula of the quasinormal mode breaks down around $R=1$. Thus we cannot say immediately 
that the axisymmetric instability mode of the black ring is found from our quasinormal mode formula. 
However $R_{f}<1$ might suggest that the unstable region for the axisymmetric perturbations exists around $R=1$. This may be related with the instability of the fat black ring
\cite{Elvang:2006dd,Figueras:2011he}

\subsection{Phase diagram}

\begin{figure}[t]
 \begin{center}
  \includegraphics[width=65mm,angle=0]{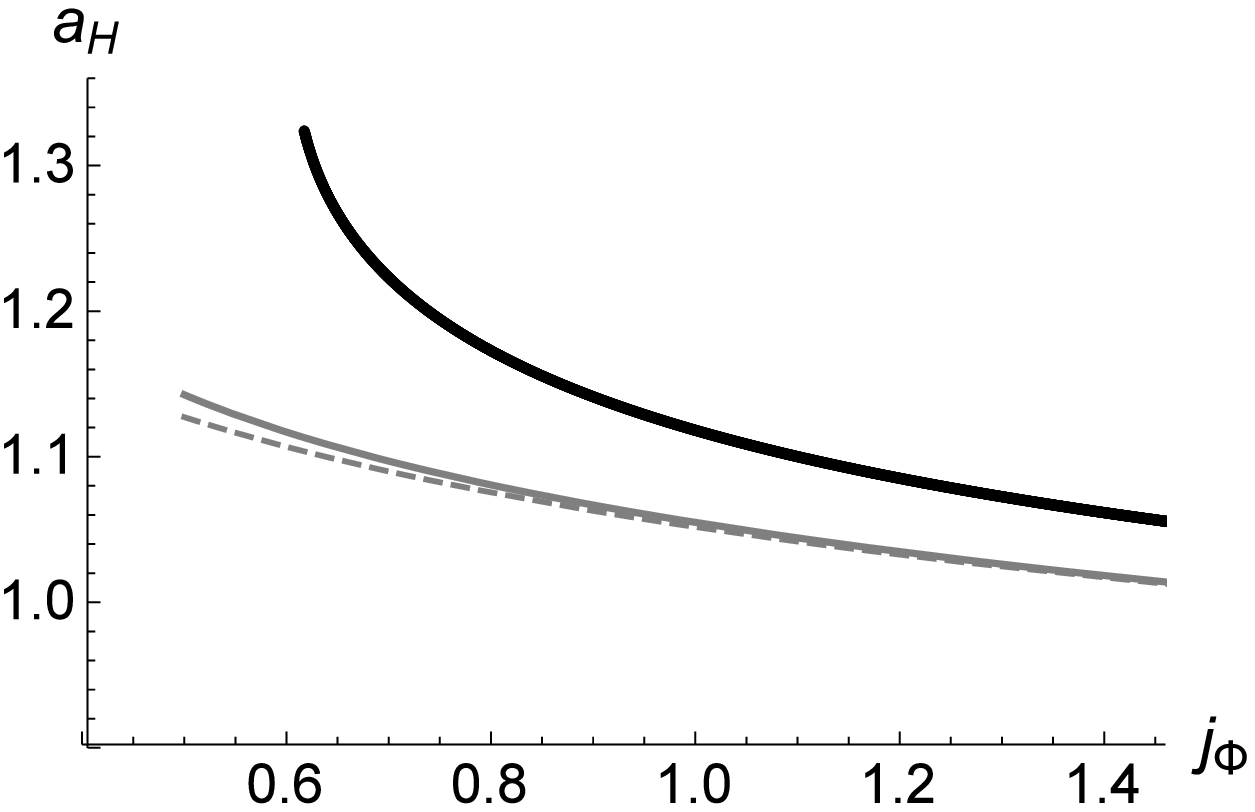}
  \hspace{5mm}
  \includegraphics[width=65mm,angle=0]{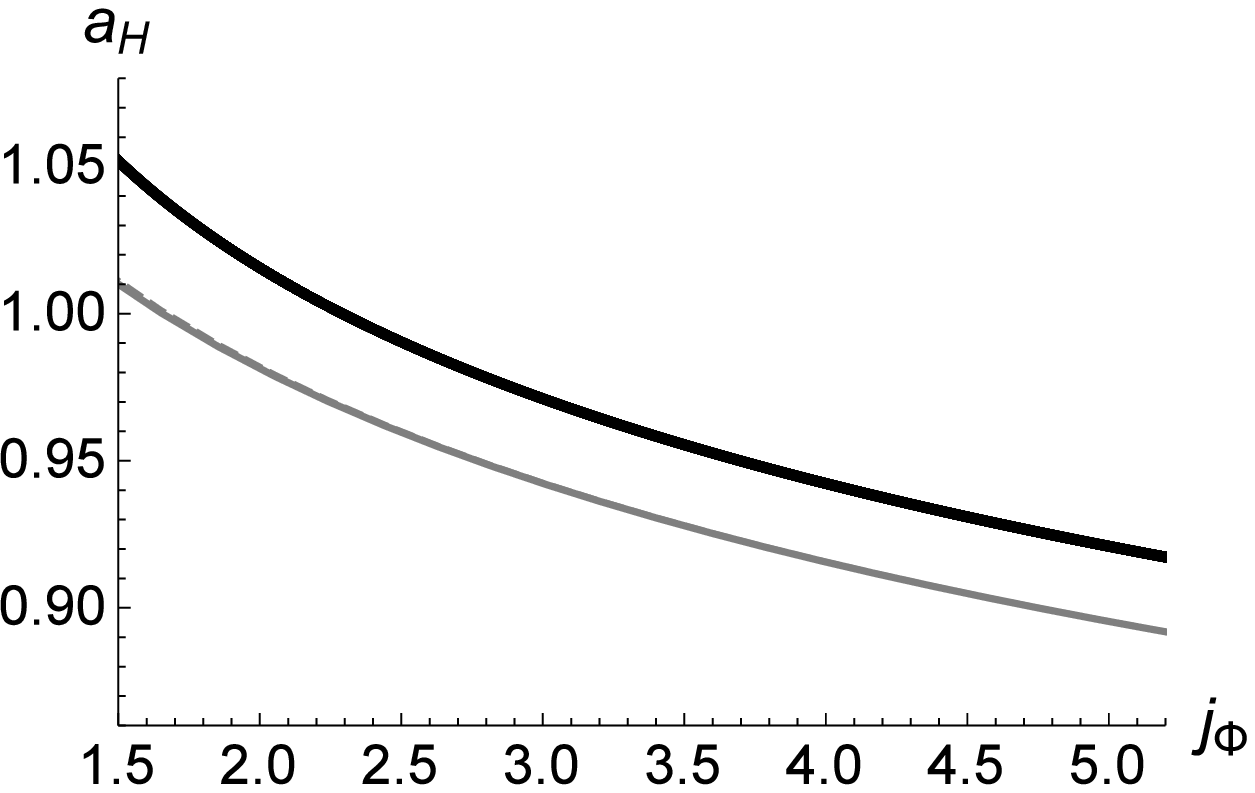}
 \end{center}
 \vspace{-5mm}
 \caption{ The $(j_{\Phi},a_{\tH})$ phase diagram of the black ring in $n=10$ is shown. The curve by black dots is the numerical evaluation 
 of the black ring solution obtained by the large $D$ expansion method. 
The gray dashed line is the leading order result by the blackfold method. The thick gray line is the result with $O(1/R^{2})$
corrections by \cite{Armas:2014bia}. The left panel shows the phase diagram in not-thin region, and right panel is for the thin 
region $R\gg r_{0}$. At the large radius $R\gg r_{0}$ these results show similar behavior as expected. The difference of values 
are within $O(1/n)$ errors in thin region.  
}
 \label{japhase}
\end{figure}

Let us draw the phase diagram of the black ring solution obtained by the large $D$ expansion method above. To do it we collect the formula 
for thermodynamic quantities of the black ring. The mass $\mathcal{M}$ and the angular momentum $\mathcal{J}_{\Phi}$ formula 
are given in eqs. (\ref{MLO}) and (\ref{JLO}). For the black ring, using the ring coordinate embedding by eq. (\ref{BR}) and the 
leading order stationary solution (\ref{LOsol}), these formula become
%
\begin{eqnarray}
\mathcal{M} = \frac{n \Omega_{n}}{8 \G} \hat{\mathcal{M}},~~
\mathcal{J}_{\Phi} = \sqrt{n}\frac{ \Omega_{n}}{8 \G} \hat{\mathcal{J}}_{\Phi},
\end{eqnarray}
%
where
%
\begin{eqnarray}
\hat{\mathcal{M}}= \int dy \frac{(R^{2}-1)e^{P(y)}}{R\sqrt{1-y^{2}}}\left(\frac{R\sqrt{1-y^{2}}}{R+y}\right)^{n},
\label{BRM}
\end{eqnarray}
%
and 
%
\begin{eqnarray}
\hat{\mathcal{J}}_{\Phi}= \int dy \frac{(R^{2}-1)^{5/2}e^{P(y)}}{R(R+y)^{2}\sqrt{1-y^{2}}}\left(\frac{R\sqrt{1-y^{2}}}{R+y}\right)^{n}.
\label{BRJ}
\end{eqnarray}
%
Note that $y=\cos{\theta}$ in the ring coordinate (\ref{ringm}). 
The formula (\ref{AreaLO}) gives the area of the black ring as
%
\begin{eqnarray}
\mathcal{A}_{\tH} = 2\pi\Omega_{n} \hat{\mathcal{A}}_{\tH},
\end{eqnarray}
%
where
%
\begin{eqnarray}
&&
\hat{\mathcal{A}}_{\tH}= \int dy \frac{\sqrt{R^{2}-1}e^{P(y)}}{\sqrt{1-y^{2}}}\left(\frac{R\sqrt{1-y^{2}}}{R+y}\right)^{n}.
\end{eqnarray}
%
$P(y)$ is given in eq. (\ref{Psol}). We set to $p_{0}=d_{0}=0$ in eq. (\ref{Psol})
\footnote{
These values of $p_{0}$ and $d_{0}$ are chosen so that the solution becomes $P(y)=O(1/R)$ at 
the large radius limit $R\gg 1$. These choices do not affect the essential feature of the phase diagram.   
}. In these formula the black ring thickness $r_{0}$ is set to unity. 
Here we consider only the leading order contributions to the mass, angular momentum and area. 
This is because it seems to be difficult to track the effect of $1/n$ corrections precisely
in $\mathcal{M}$, $\mathcal{J}_{\Phi}$ and $\mathcal{A}_{\tH}$ due to a term in integrands with a power of $n$. 
At large $n$ the integrations of such terms can be very large, and it is not 
clear how to control the size of such integrations in $1/n$ expansion. In this paper, we take into account only the leading order contributions 
to observe general feature, not detail numerical values, of the phase diagram of the large $D$ black ring. 
Hence the phase diagrams for the large $D$ black ring shown below have $O(1/n)$ errors. 
On the other hand the temperature and horizon angular 
velocity do not have such troublesome terms in their definitions, so we include $1/n$ corrections. The temperature $T_{\tH}$ of the black ring is
%
\begin{eqnarray}
T_{\tH} \equiv \frac{\kappa}{2\pi} = \frac{n}{4\pi}\frac{\sqrt{R^{2}-1}}{R}\Biggl[
 1 - \frac{1}{2n}
\Biggr].
\end{eqnarray}
%
The horizon angular velocity of the black ring is
%
\begin{eqnarray}
\Omega_{H}=\frac{\hat{\Omega}_{\tH}}{\sqrt{n}}=\frac{1}{\sqrt{n}}\frac{\sqrt{R^{2}-1}}{R^{2}} \Biggl[
1-\frac{1}{n}\frac{R^{2}-2\log{(R-R^{-1})}}{2R^{2}}
\Biggr].
\end{eqnarray}
%
To draw the phase diagram we define reduced quantities by the mass as following
%
\begin{eqnarray}
j_{\Phi}^{n+1}=c_{j}\frac{\mathcal{J}_{\Phi}^{n+1}}{\G \mathcal{M}^{n+2}},~~
a_{\tH}^{n+1}=c_{a}\frac{\mathcal{A}^{n+1}_{\tH}}{(\G \mathcal{M})^{n+2}},
\end{eqnarray}
%
and
%
\begin{eqnarray}
t_{\tH}=c_{T}T_{\tH}(\G \mathcal{M})^{1/(n+1)},~~
\omega_{\tH} = c_{\omega}\Omega_{\tH}(\G \mathcal{M})^{1/(n+1)}.
\end{eqnarray}
%
The normalization numerical coefficients are taken from \cite{Emparan:2007wm} as
%
\begin{eqnarray}
c_{j}=\frac{\Omega_{n+1}}{2^{n+5}}\frac{(n+2)^{n+2}}{(n+1)^{(n+1)/2}},~~
c_{a}=\frac{\Omega_{n+1}}{2(16\pi)^{n+1}}\frac{n^{(n+1)/2}(n+2)^{n+2}}{(n+1)^{(n+1)/2}}, 
\end{eqnarray}
%
and
%
\begin{eqnarray}
c_{t}=\frac{4\pi\sqrt{n+1}}{\sqrt{n}}\left( \frac{(n+2)\Omega_{n+1}}{2} \right)^{-1/(n+1)},
c_{\omega} = \sqrt{n+1} \left( \frac{(n+2)\Omega_{n+1}}{16} \right)^{-1/(n+1)}.
\end{eqnarray}
%
Then we obtain
%
\begin{eqnarray}
j_{\Phi}^{n+1}=\frac{1}{2^{n+5}}\frac{\Omega_{n+1}}{\Omega_{n}}\left( 1+\frac{2}{n} \right)^{n+2}\left( 1+\frac{1}{n} \right)^{-(n+1)/2}
\frac{\hat{\mathcal{J}}_{\Phi}^{n+1}}{\hat{\mathcal{M}}^{n+2}},
\label{phasej}
\end{eqnarray}
%
%
\begin{eqnarray}
a_{\tH}^{n+1}=\frac{1}{2^{n}}\frac{\Omega_{n+1}}{\Omega_{n}}\left( 1+\frac{2}{n} \right)^{n+2}\left( 1+\frac{1}{n} \right)^{-(n+1)/2}
\frac{\hat{\mathcal{A}}_{\tH}^{n+1}}{\hat{\mathcal{M}}^{n+2}},
\label{phasea}
\end{eqnarray}
%
%
\begin{eqnarray}
t_{\tH}=2^{n/(n+1)}\left( 1+\frac{1}{n} \right)^{1/2}\left( \frac{\Omega_{n}}{\Omega_{n+1}} \right)^{1/(n+1)}\left( 1+\frac{2}{n} \right)^{-1/(n+1)}
\kappa\hat{\mathcal{M}}^{1/(n+1)},
\label{phaset}
\end{eqnarray}
%
and
%
\begin{eqnarray}
\omega_{\tH}=2^{2/(n+1)}\left( 1+\frac{1}{n} \right)^{1/2}\left( \frac{\Omega_{n}}{\Omega_{n+1}} \right)^{1/(n+1)}\left( 1+\frac{2}{n} \right)^{-1/(n+1)}
\hat{\Omega}_{\tH}\hat{\mathcal{M}}^{1/(n+1)}.
\label{phasew}
\end{eqnarray}
%
We evaluate $j_{\Phi}$, $a_{\tH}$, $t_{\tH}$ and $\omega_{\tH}$ numerically for the large $D$ black ring solution by using eqs. (\ref{BRM}) and (\ref{BRJ}) with $P(y)$
given in eq. (\ref{Psol}). As mentioned above, we set to $p_{0}=d_{0}=0$. 
Figure \ref{japhase} shows the phase diagram of $(j_{\Phi},a_{\tH})$ for the black ring solution by the blackfold and large $D$ expansion method 
from $R/r_{0}=1.1$ to $R/{r_{0}}=20$ in $n=10$.
$R$ and $r_{0}$ are a ring radius and ring thickness of the black ring.  
%
%
The curve by black dots is the result by numerical evaluations of eqs. (\ref{phasej}) and (\ref{phasea}) for our large $D$ black ring solution. 
The gray dashed line is the leading order result by the blackfold method 
\cite{Emparan:2007wm}. The thick gray line is $O(1/R^{2})$  correction of the blackfold \cite{Armas:2014bia}. As expected we can see that 
these results show similar behavior at the large radius region $R\gg 1$. Two results by the blackfold and large $D$ expansion method 
seems to have a difference by the constant offset at $R\gg 1$. This difference can be understood as $O(1/n)$ errors in the large $D$ expansion
method. By taking the large radius limit $R\gg 1$, from $P(y)=O(1/R)$, we find that
%
\begin{eqnarray}
\mathcal{M}=\frac{n\Omega_{n+1}}{8\G}Rr_{0}^{n}\left( 1+O(n^{-1},R^{-1}) \right),~
\mathcal{J}_{\Phi}=\frac{\sqrt{n}\Omega_{n+1}}{8\G}R^{2}r_{0}^{n}\left( 1+O(n^{-1},R^{-1}) \right),
\end{eqnarray}
%
and
%
\begin{eqnarray}
\mathcal{A}_{\tH} = 2\pi\Omega_{n+1}Rr_{0}^{n+1}\left( 1+O(n^{-1},R^{-1}) \right),
\end{eqnarray}
%
where we restored the black ring thickness $r_{0}$. On the other hand the leading order results by the blackfold \cite{Emparan:2007wm} gives
%
\begin{eqnarray}
\mathcal{M}=\frac{(n+2)\Omega_{n+1}}{8\G}Rr_{0}^{n}\left( 1+O(R^{-1}) \right),~
\mathcal{J}_{\Phi}=\frac{\sqrt{n+1}\Omega_{n+1}}{8\G}R^{2}r_{0}^{n}\left( 1+O(R^{-1}) \right),
\label{BFLOmj}
\end{eqnarray}
%
and
%
\begin{eqnarray}
\mathcal{A}_{\tH} = 2\pi\sqrt{\frac{n+1}{n}}\Omega_{n+1}Rr_{0}^{n+1}\left( 1+O(R^{-1}) \right).
\label{BFLOa}
\end{eqnarray}
%
These results coincide at the leading order in $1/n$ expansion. $1/n$ corrections in eqs. (\ref{BFLOmj}) and (\ref{BFLOa})
gives smaller $a_{\tH}$ and $j_{\Phi}$ than the leading order results in $1/n$ expansion. So $1/n$ correction would reduce 
the difference seen in Figure \ref{japhase}.
This $O(1/n)$ error would give the difference of the onset at $R\gg 1$ seen in Figure \ref{japhase}. It might be difficult
to observe $O(1/R^{2})$ corrections in the results by the large $D$ expansion method. This is because the definitions of thermodynamic quantities
contain the $y$-integrations of $R^{n}/(R+y)^{n}$. The blackfold by the $1/R$ expansion and the large $D$ expansion method by $1/n$ expansion 
do not give same results for such integrations in higher order corrections of $O(n^{-1},R^{-2})$.

%
\begin{figure}[t]
 \begin{center}
  \includegraphics[width=65mm,angle=0]{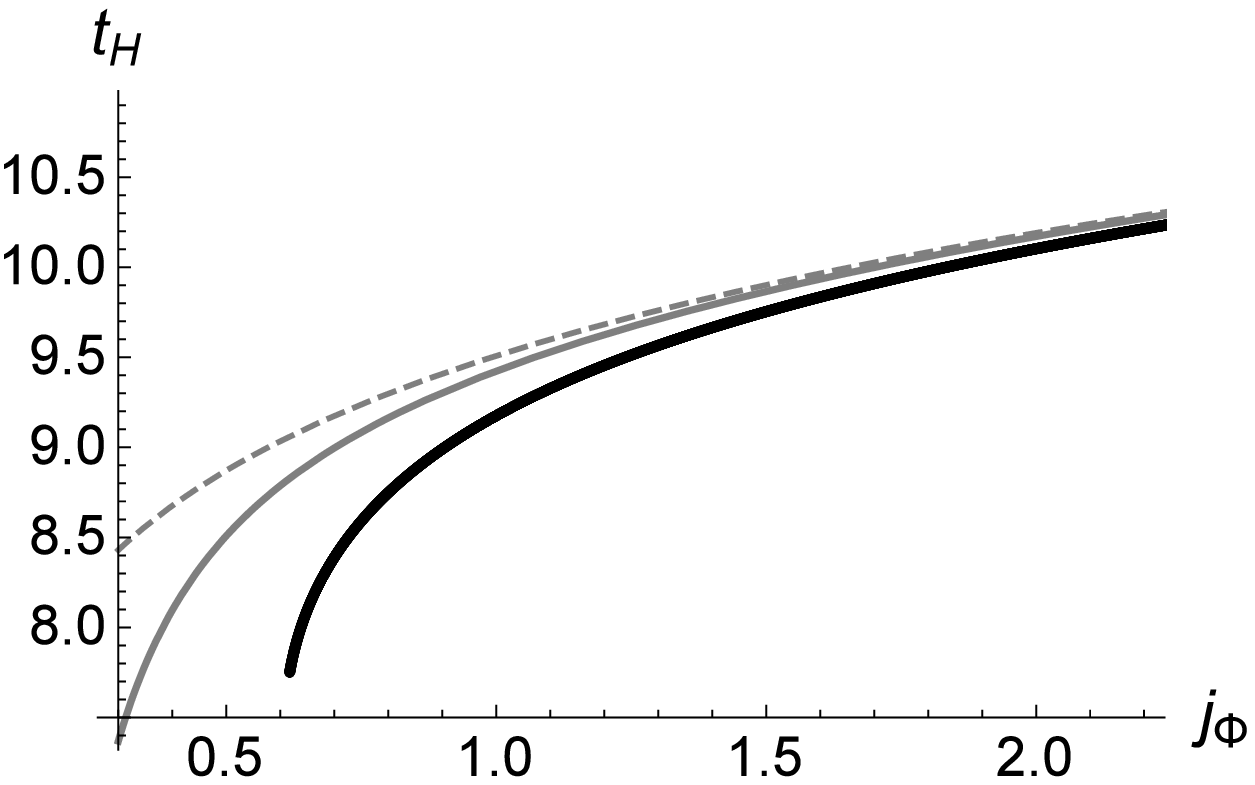}
  \hspace{5mm}
  \includegraphics[width=65mm,angle=0]{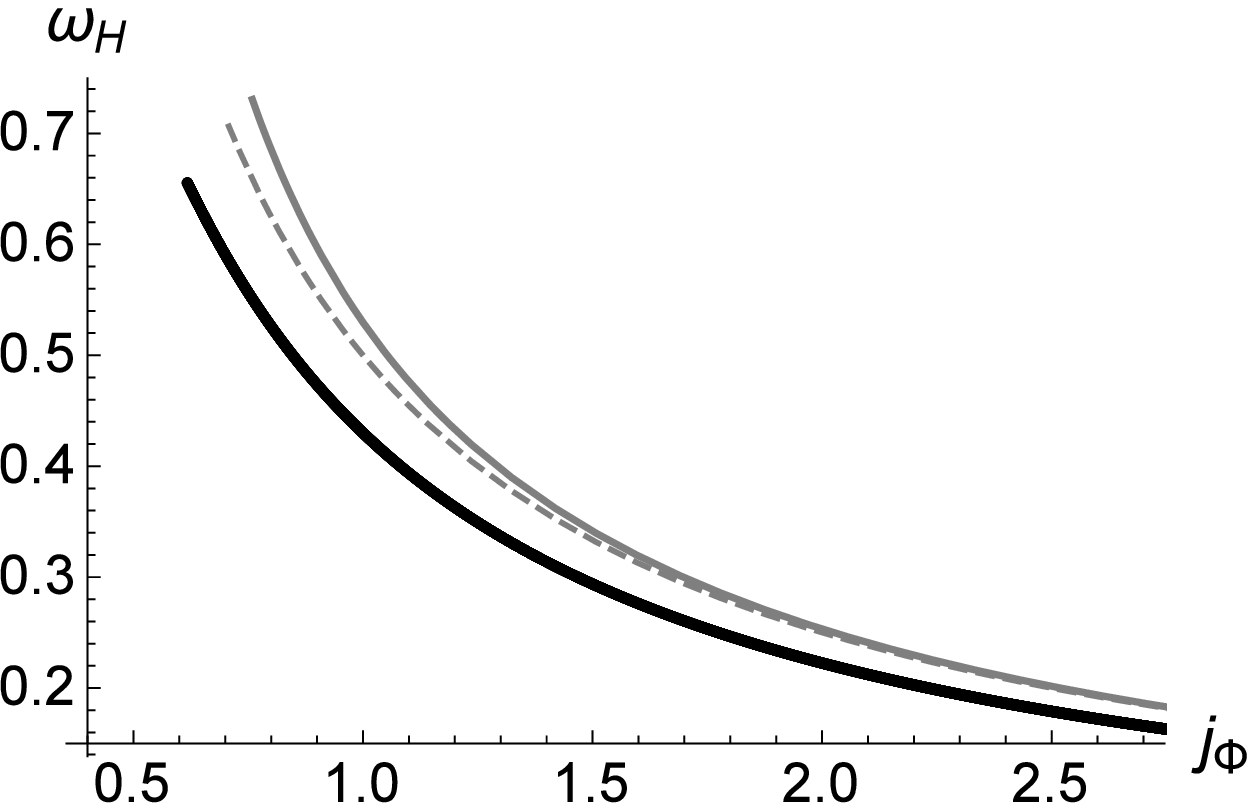}
 \end{center}
 \vspace{-5mm}
 \caption{ The $(j_{\Phi},t_{\tH})$ (left) and $(j_{\Phi},\omega_{\tH})$ (right) phase diagram in $n=10$ are shown. The gray dashed and 
 thick line corresponds to the leading order and $O(1/R^{2})$ corrected results of the blackfold method. The curve by black dots is the
numerical evaluations for the large $D$ black ring solution.  
}
 \label{jtwphase}
\end{figure}
%
Figure \ref{jtwphase} are plots of the phase diagram of $(j_{\Phi},t_{\tH})$ and $(j_{\Phi},\omega_{\tH})$ for the black ring
by the blackfold (gray dashed and thick lines) and the large $D$ expansion method (curve by black dots). 
In these diagrams we see that results by the blackfold and the large $D$ expansion method show similar values and behaviors 
at the large radius region $R\gg 1$. These results show different behavior in not-thin region $j_{\Phi}\sim 1$. This might be because our
results contain $O(1/n)$ errors which become larger at fat ring region, and the blackfold method also has not-small errors there
\footnote{
In not so much higher dimension, it was observed that the blackfold method has a good accuracy even in not thin region $j_{\Phi}\sim 1$ if we include higher order corrections \cite{Armas:2014bia}. 
}.

\section{Summary} \label{4}

We constructed the effective theory for the slowly rotating large $D$ black holes. The slow rotation is defined by $O(1/\sqrt{D})$ horizon angular velocity at large $D$. This solution class of large $D$ black holes contains the slowly rotating Myers-Perry black hole,
slowly boosted black string and black ring solution. The black ring should be slowly rotating at large $D$ since the tension effect
which determines the horizon angular velocity is small. So our analysis of the black ring in this paper by the large $D$ effective theory is not restricted one. 
We solved the effective equations and found the black ring solution 
analytically. Furthermore, by perturbation analysis, the quasinormal mode condition of the black ring was obtained. 
The quasinormal mode says that the thin black ring has the Gregory-Laflamme type instability in the non-axisymmetric perturbations as
found numerically. The black ring solution obtained in this paper describes not only thin black ring, but also not-thin black ring. 
For not-thin black ring, we found that the black ring becomes stable against non-axisymmetric perturbations. We gave some discussions
on the endpoint of the non-axisymmetric instabilities of the black ring. At large $D$ the gravitational wave emission becomes
much less efficient by $O(e^{-D/2})$ than the dynamical instability, and the instability would not lead to the fragmentation of the
horizon in enough higher dimensions. This fact suggests the existence of a metastable solution, non-uniform black ring, as the endpoint of
the black ring instability at large $D$. We also studied $1/D$ corrections to the quasinormal modes and phase diagram. 

In this paper we considered the construction of the black ring solution by the large $D$ expansion method as the first step to reveal the 
variety of higher dimensional black holes. So we have some natural extensions of our current work. One is to consider 
the $O(1)$ horizon angular velocity solution. In \cite{Suzuki:2015iha} the effective theory for the stationary solution
with $O(1)$ horizon angular velocity was constructed, but they did not include the time dependent deformations. 
It is possible to include the time dependent deformations in the work \cite{Suzuki:2015iha} by the same way in this paper. 
The non-liner dynamical deformations of the singly rotating Myers-Perry black hole can be described by such framework. 
The singly rotating Myers-Perry black hole is known to have the unstable modes, and it is interesting to study 
the instability mode and stationary deformed solution associated with the zero mode, so called bumpy black hole, as the endpoint 
of the instability by the large $D$ effective theory. 
The second possible extension is to investigate the non-linear evolution of the instability of the black ring. The effective equations obtained in this paper 
describe such evolutions. It is interesting to solve the effective equations numerically and study the endpoint of the black ring instability. 
Our effective equations have much simpler form than original Einstein equations, and they are expected to be much more tractable numerically. 
In that analysis our conjecture for the existence of the non-uniform black ring can be also studied. 
The third is to include further angular momentum to the effective equations. The effective equations in this paper 
describe the dynamical black hole with essentially only one angular momentum. By adding further number of angular momenta to the effective equations, we can describe 
much richer dynamics of the black holes, and it will be found that the variety of the 
black hole horizon becomes much richer than one of the singly rotating black hole. These work would give new interesting insights to
the higher dimensional black hole physics.

\section*{Acknowledgments}

The author was grateful to Roberto Emparan for useful discussions and comments on the draft.  
This work was supported by JSPS Grant-in-Aid for Scientific Research No.26-3387.

\appendix 

\section{Other solutions} \label{B}

In this appendix we study other embeddings of the solution. In the asymptotic region $\sR\gg 1$ 
the leading order metric on $\sR=\text{const.}$ surface has the following form
%
\begin{eqnarray}
ds^{2} |_{\sR=\text{const.}} =-dv^{2} +r^{2}\left( dz^{2}+ G(z)^{2}d\Phi^{2} +H(z)^{2}d\Omega^{2}_{n} \right) +O(\sR^{-1}).
\label{LOm}
\end{eqnarray}
%
$H(z)$ should satisfy the equation given by
%
\begin{eqnarray}
1-H'(z)^{2}+H(z)H''(z)=0.
\label{HeqApp}
\end{eqnarray}
%
In the following this metric is embedded into a flat background in spherical coordinates. Such solutions 
describes the slowly rotating Myers-Perry black hole and slowly boosted black string solution.

\subsection{Slowly rotating Myers-Perry black hole}

The first spherical coordinate is the spherical coordinate of $D=n+4$ dimensional flat spacetime. The metric
in the spherical coordinate is
%
\begin{eqnarray}
ds^{2}=-dt^{2}+dr^{2}+r^{2}(dz^{2}+\sin^{2}{z}~d\Phi^{2}+\cos^{2}{z}~d\Omega^{2}_{n}). 
\label{SPH1}
\end{eqnarray}
%
The embedding of the leading order metric is given by $r=r_{0}$ in this flat background metric. 
Then the embedding says that $H(z)$ and $G(z)$ are identified by
%
\begin{eqnarray}
H(z)=\cos{z},~~
G(z)=\sin{z}. 
\end{eqnarray}
%
Note that this $H(z)$ actually satisfies eq. (\ref{HeqApp}). The surface gravity of this spherical coordinate embedding is
%
\begin{eqnarray}
\kappa = n \hat{\kappa} 
=\frac{n}{2}.
\end{eqnarray}
%
$r=\text{const.}$ surfaces in the metric (\ref{SPH1}) have the topology of $S^{n+2}$. 
So the topology of the black hole horizon is $S^{n+2}$ in this embedding. 

\paragraph{Effective equations}

The effective equations (\ref{Deq1}), (\ref{Deq2}) and (\ref{Deq3}) in the spherical coordinate embedding are
%
\begin{eqnarray}
\pv p_{v} +\pz p_{v}\tan{z} -\frac{\pp^{2}p_{v}}{\sin^{2}{z}}+\frac{\pp p_{\phi}}{\sin^{2}{z}}-p_{z}\tan{z}=0,
\label{Deq1SPH}
\end{eqnarray}
%
%
\begin{eqnarray}
\pv p_{\phi}+\pz p_{\phi}\tan{z} -\frac{\pp^{2}p_{\phi}}{\sin^{2}{z}}+\pp p_{v} +\frac{1}{\sin^{2}{z}}\pp \Biggl[ \frac{p_{\phi}^{2}}{p_{v}} \Biggr]
-2p_{\phi} -\frac{p_{\phi}p_{z}}{p_{v}}\tan{z} =0,
\label{Deq2SPH}
\end{eqnarray}
%
and
%
\begin{eqnarray}
&&~~
\pv p_{z}+\pz p_{z} \tan{z} -\frac{\pp^{2}p_{z}}{\sin^{2}{z}} +\pz p_{v} +\frac{1}{\sin^{2}{z}}\pz \Biggl[ \frac{p_{z}p_{\phi}}{p_{v}} \Biggr] \notag \\
&&~~~~~~~~~~~~~~~~
-\frac{p_{\phi}^{2}\cot{z}+p_{z}^{2}\tan{z}\sin^{2}{z}}{p_{v}\sin^{2}{z}}
+\frac{2\cot{z}}{\sin^{2}{z}}\pp p_{\phi} -\frac{\cos{2z}}{\cos^{2}{z}}p_{z} =0.
\label{Deq3SPH}
\end{eqnarray}
%

\paragraph{Stationary solution}

We solve eqs. (\ref{Deq1SPH}), (\ref{Deq2SPH}) and (\ref{Deq3SPH}) for the stationary solution. The stationary solution ansatz assumes that 
$\pv$ and $\pp$ are the Killing vectors. Thus we assume
%
\begin{eqnarray}
p_{v}=e^{P(z)},~~p_{\phi}=p_{\phi}(z),~~p_{z}=p_{z}(z).
\end{eqnarray}
%
$p_{a}(z)$ are given by $p_{v}(z)$ as derived in eq. (\ref{psol}) with the integration constant $\hat{\Omega}_{\tH}$. 
The equation for $P(z)$ given in eq. (\ref{PeqS}) in the spherical coordinate embedding is
%
\begin{eqnarray}
P''(z)+P'(z)\tan{z}-\hat{a}^{2}\cos^{2}{z}=0.
\end{eqnarray}
%
Here we rename $\hat{\Omega}_{\tH}$ in eq. (\ref{PeqS}) by $\hat{a}$ just for usefulness. 
The solution of this equation is
%
\begin{eqnarray}
P(z) = p_{0}+d_{0}\sin{z}-\frac{1}{2}\hat{a}^{2}\cos^{2}{z}. 
\end{eqnarray}
%
Integration constants, $p_{0}$ and $d_{0}$, describe trivial deformations of the solution, so we set to $p_{0}=d_{0}=0$. 
Then, using eq. (\ref{psol}), we find that the stationary solutions of eqs. (\ref{Deq1SPH}), (\ref{Deq2SPH}) and (\ref{Deq3SPH}) are
%
\begin{eqnarray}
p_{v}(z)=e^{-\hat{a}^{2}\cos^{2}{z}/2},~~p_{\phi}(z)=\hat{a}p_{v}(z)\sin^{2}{z},~~
p_{z}(z)=p_{v}'(z).
\end{eqnarray}
%
This stationary solution describes the $D=n+4$ dimensional slowly rotating Myers-Perry black hole \cite{Myers:1986un} with $a=\hat{a}/\sqrt{n}$.
This can be confirmed directly by performing a coordinate transformation from the Myers-Perry black hole. As we can see below, 
quasinormal modes of the stationary solution also coincide to one of the Myers-Perry black hole obtained found in \cite{Suzuki:2015iha}. 

\paragraph{Quasinormal modes}

By perturbing the equations (\ref{Deq1SPH}), (\ref{Deq2SPH}) and (\ref{Deq3SPH}) around the stationary solution, we obtain
quasinormal mode frequencies. The perturbations are
%
\begin{eqnarray}
&&
p_{v}(v,z,\phi) =e^{-a^{2}\cos^{2}{z}/2} \left( 1+\epsilon F_{v}(z)e^{-i\omega v}e^{im\phi} \right), \notag \\
&&
p_{\phi}(v,z,\phi) = a\sin^{2}{z}e^{-a^{2}\cos^{2}{z}/2}\left( 1+\epsilon F_{\phi}(z)e^{-i\omega v}e^{im\phi} \right), \notag \\
&&
p_{z}(v,z,\phi) = a^{2}\cos{z}\sin{z}e^{-a^{2}\cos^{2}{z}/2}\left( 1+\epsilon F_{z}(z)e^{-i\omega v}e^{im\phi} \right).
\label{pSCH}
\end{eqnarray}
%
We impose the boundary condition on the perturbations at $z=0$ as
\footnote{
This boundary condition was also used in \cite{Suzuki:2015iha}.
}
%
\begin{eqnarray}
F_{v}(z) \propto z^{\ell} \left( 1+O(z) \right).
\label{bcpSCH}
\end{eqnarray}
%
Then the frequency is discretized by quantum numbers, $\ell$ and $m$.  
In the following we assume $m=O(n^{-1})$. This assumption can be understood by observing the spherical harmonics. 
At large $D$ the spherical harmonics on $S^{n+1}$, $\mathbb{Y}_{j\ell m_{\Phi}}(z)$, in the metric (\ref{SPH1}) 
is reduced to \cite{Suzuki:2015iha}
%
\begin{eqnarray}
\mathbb{Y}_{j\ell m_{\Phi}}(z) \propto  e^{im_{\Phi}\Phi}\left( \sin{z} \right)^{|m_{\Phi}|+2k_{S}}\cos^{j}{z}
\label{SPHH}
\end{eqnarray}
%
where $j$ is the quantum number on $S^{n}$. $k_{S}$ is the quantum number along $z$-direction. Comparing eqs. (\ref{bcpSCH}) and
(\ref{SPHH}), we see that $\ell$ is given by
%
\begin{eqnarray}
\ell = |m_{\Phi}|+2k_{S}. 
\end{eqnarray}
%
In this paper we do not consider the excitation on $S^{n}$, so our boundary condition corresponds to one of $j=0$ in \cite{Suzuki:2015iha}. 
Note that, since we rescaled $\phi$ coordinate by eq. (\ref{phiRe}), the quantum number $m_{\Phi}$ associated with $\partial_{\Phi}$ 
is related with the quantum number $m$ of $\pp$ by $m_{\Phi}=\sqrt{n}m$. 
Thus the spherical harmonics has $\sqrt{n}m$ dependence, not $m$ dependence. If we assume $m=O(1)$, the spherical harmonics is dominated by the quantum number $m$ as
%
\begin{eqnarray}
\mathbb{Y}_{j\ell m_{\Phi}}(z) \propto (\sin{z})^{\sqrt{n}m}.
\end{eqnarray}
%
The analysis of this function becomes involved since the $z$-derivative can be very large and diverging in $n$. To avoid this complexity we assume $m=O(1/n)$. 
Then the perturbations can satisfy the boundary condition (\ref{bcpSCH}) without diverging terms in $n$. 
Thus, in the perturbation (\ref{pSCH}), we introduce new $O(1)$ quantity $\bar{m}$ by
%
\begin{eqnarray}
m = \frac{\bar{m}}{n}. 
\end{eqnarray}
%
Then the quasinormal mode frequencies are written by $\ell$, $\bar{m}$ and $\hat{a}$. 
We can extend the all calculations up to $O(1/n)$ by including $1/n$ corrections. The obtained quasinormal modes up to $O(1/n)$ is
%
\begin{eqnarray}
\omega_{\pm} = \omega_{\pm}^{(0)}+\frac{\omega_{\pm}^{(1)}}{n}+O(n^{-2}),
\end{eqnarray}
%
where
%
\begin{eqnarray}
\omega_{\pm}^{(0)}=\pm\sqrt{\ell-1}-i(\ell-1)
\end{eqnarray}
%
and
%
\begin{eqnarray}
&&
\omega_{\pm}^{(1)} = \pm\Biggl(
\frac{\sqrt{\ell-1}(3\ell-4)}{2} -\frac{\hat{a}^{2}(2\ell^{2}-9\ell+8)}{2\ell\sqrt{\ell-1}}
+\frac{i\hat{a}\bar{m}\sqrt{\ell-1}}{\ell}
\Biggr) \notag \\
&&~~~~~~~~~~~
-\Biggl(
i(\ell-1)(\ell-2)-\frac{i\hat{a}^{2}(\ell^{2}-5\ell+8)}{2\ell} -\frac{\hat{a}\bar{m}(\ell-1)}{\ell}
\Biggr).
\end{eqnarray}
%
Actually this quasinormal mode reproduces the quasinormal mode of the Schwarzschild black hole up to $1/n$ 
corrections for $\hat{a}=0$ \cite{Emparan:2014aba} and one of Myers-Perry black hole \cite{Suzuki:2015iha} with $a=\hat{a}/\sqrt{n}$ and 
$m=\bar{m}/\sqrt{n}$.

\subsection{Slowly boosted black string}

We consider the embedding in another spherical coordinate, which is the coordinate of the spacetime with one compact direction. 
The $D=n+4$ dimensional spacetime with one compact direction has
the following metric in the spherical coordinate 
%
\begin{eqnarray}
ds^{2}=-dt^{2}+d\Phi^{2}+dr^{2}+r^{2}\left( dz^{2}+\sin^{2}{z}~d\Omega^{2}_{n} \right). 
\label{fBS}
\end{eqnarray}
%
$\Phi$ is a coordinate of the compact direction. 
The embedding  $r=r_{0}$ of the leading order metric (\ref{LOm}) into eq. (\ref{fBS}) gives following identifications
%
\begin{eqnarray}
G(z)=1,~~H(z)=\sin{z},
\end{eqnarray}
%
where we set to $r_{0}=1$. 
The embedded solution has $S^{1}\times S^{n+1}$ horizon topology in the spacetime with one compact dimension. So the solution of the effective equations describe 
non-linear dynamical deformations of the black string.

\paragraph{Effective equations}

The effective equations in this embedding become
%
\begin{eqnarray}
\pv p_{v}-\cot{z}~\pz p_{v}-\pp^{2}p_{v}+\pp p_{\phi}+p_{z}\cot{z}=0,
\end{eqnarray}
%
%
\begin{eqnarray}
\pv p_{\phi}-\cot{z}~\pz p_{\phi}-\pp^{2}p_{\phi}-\pp p_{v} +\pp\Biggl[ \frac{p_{\phi}^{2}}{p_{v}} \Biggr]-\cot{z}\frac{p_{z}p_{\phi}}{p_{v}}=0,
\end{eqnarray}
%
and
%
\begin{eqnarray}
\pv p_{z}-\cot{z}~\pz p_{z}-\pp^{2}p_{z}+\pz p_{v} +\partial_{z}\Biggl[ \frac{p_{z}p_{\phi}}{p_{v}} \Biggr] -\cot{z}\frac{p_{z}^{2}}{p_{v}} 
-\frac{\cos{2{z}}}{\sin^{2}{z}}p_{z}=0.
\end{eqnarray}
%

\paragraph{Stationary solutions}
We have two stationary solutions of the effective equations. One is the boosted black string. 
The ansatz for the boosted black string is 
%
\begin{eqnarray}
p_{v}(z)=e^{P(z)},~~p_{\phi}=p_{\phi}(z),~~p_{z}=p_{z}(z).
\end{eqnarray}
%
Then the effective equations give the solution 
%
\begin{eqnarray}
p_{v}(z)=e^{p_{0}+d_{0}\cos{z}},~~p_{\phi}=\hat{\sigma} p_{v}(z),~~p_{z}=p_{v}'(z).
\label{boostsolapp}
\end{eqnarray}
%
$\hat{\sigma}$ is an integration constant, and it describes the boost parameter of the boosted black string with the
boost parameter
%
\begin{eqnarray}
\sinh{\alpha}=\frac{\hat{\sigma}}{\sqrt{n}},
\end{eqnarray}
%
where the boost transformation of the black string and definition of $\alpha$ are given in eq. (\ref{boosttrans}). 
The integration constants $p_{0}$
and $d_{0}$ are horizon size and position of the black string origin respectively. Thus we can set to $p_{0}=0$ and $d_{0}=0$. 

Another stationary solution is the non-uniform black string. This solution is inhomogeneous along $\Phi$ direction. 
So $\partial_{\Phi}$ is not the Killing vector. The ansatz for the non-uniform black string is
%
\begin{eqnarray}
p_{v}=p_{v}(\phi),~~p_{\phi}=p_{\phi}(\phi),~~p_{z}=0.
\end{eqnarray}
%
The condition $p_{z}=0$ reflects the fact that the Gregory-Laflamme mode of the black string exists only in the 
S-wave sector. This ansatz gives the equation for the non-uniform black string at large $D$ \cite{Suzuki:2015axa} as
%
\begin{eqnarray}
p_{\phi} = \pp p_{v} +P_{\phi}
\end{eqnarray}
%
and
%
\begin{eqnarray}
\pp^{3}p_{v}+\pp p_{v}-\frac{2\pp p_{v}\pp^{2}p_{v}}{p_{v}}+\frac{(\pp p_{v})^{3}}{p_{v}^{2}}=0.
\label{pveq}
\end{eqnarray}
%
$P_{\phi}$ is an integration constant describing the momentum along $\phi$ direction, and we set to $P_{\phi}=0$ in eq. (\ref{pveq}). 
To solve this equation we need the numerical treatment. 
In this paper we consider only the boosted black string solution.

\paragraph{Quasinormal modes}

Considering the perturbations around the stationary solution (\ref{boostsolapp})
%
\begin{eqnarray}
&&
p_{v}(v,z,\phi) = 1+\epsilon F_{v}(z)e^{-i\omega v}e^{i\hat{k}\phi} , \notag \\
&&
p_{\phi}(v,z,\phi) = \hat{\sigma} \left( 1+\epsilon F_{\phi}(z)e^{-i\omega v}e^{i\hat{k}\phi} \right), \notag \\
&&
p_{z}(v,z,\phi) = \epsilon F_{z}(z)e^{-i\omega v}e^{i\hat{k}\phi}, 
\label{pBS}
\end{eqnarray}
%
and boundary conditions at $\cos{z}=0$ given by
%
\begin{eqnarray}
F_{v}(z) \propto (\cos{z})^{\ell}\left( 1+O(\cos{z}) \right),
\end{eqnarray}
%
we obtain the quasinormal mode condition as
%
\begin{eqnarray}
&&
\frac{1}{\hat{k}+\hat{\sigma}^{2}\hat{k}+\hat{\sigma}(i\ell +i\hat{k}^{2}+\omega)}\Bigl[
\omega^{3}+(-2i+3i\ell+3\hat{\sigma} \hat{k} +3i\hat{k}^{2})\omega^{2} \notag \\
&&~~~
+(-3\ell^{2}+\ell(3+6i\hat{\sigma} \hat{k} -6\hat{k}^{2}) +\hat{k}(5\hat{k}+3\hat{\sigma}^{2}\hat{k} 
-3\hat{k}^{2} +2i\hat{\sigma}(3\hat{k}^{2}-2)) )\omega \notag \\
&&~~~
-i(\ell^{3}+\ell^{2}(-1-3i\hat{\sigma} \hat{k}+3\hat{k}^{2}) +k\ell(3i\hat{\sigma} -4\hat{k}-3\hat{\sigma}^{2}\hat{k}
-6i\hat{\sigma} \hat{k}+3\hat{k}^{2} ) \notag \\
&&~~~~
+\hat{k}^{2}(2+i\hat{\sigma}^{3}\hat{k} -3\hat{k}^{2} +\hat{k}^{4} +\hat{\sigma}^{2}(2-3\hat{k}^{2})-i\hat{\sigma} \hat{k} (3\hat{k}^{2}-5))
\Bigr]=0.
\label{QNMBS}
\end{eqnarray}
%
This condition can be solved for $\ell=0$ mode in a simple form by
%
\begin{eqnarray}
\omega^{(\ell=0)}_{\pm} = \hat{\sigma}\hat{k} \pm i\hat{k}(1\mp \hat{k}).
\end{eqnarray}
%
This is the quasinormal modes of the S-wave sector of the scalar type gravitational perturbation of the boosted black string. 
$\omega^{(\ell=0)}_{+}$ shows the Gregory-Laflamme instability \cite{Gregory:1993vy} for $\hat{k}<1$. For $\hat{k}=0$ mode the quasinormal mode condition gives
%
\begin{eqnarray}
\omega^{(\hat{k}=0)}_{\pm} = \pm\sqrt{\ell-1}-i(\ell-1). 
\end{eqnarray}
%
Although, for general modes $\ell\neq 0$ and $\hat{k}\neq 0$, we cannot obtain a simple solution of eq. (\ref{QNMBS}), 
it can be seen that the instability mode exists only in the S-wave ($\ell=0$) sector. 
This is consistent with the result in \cite{Kudoh:2006bp} that the black string is unstable
only for the S-wave sector.

We can also find $1/n$ corrections to the quasinormal modes. For the S-wave sector, the quasinormal modes are
%
\begin{eqnarray}
&&
\omega^{(\ell=0)}_{\pm} =\hat{\sigma} \hat{k} \pm i\hat{k}(1\mp \hat{k}) \notag \\
&&~~~~
+\frac{\hat{k}}{2n}\Bigl[
i(\mp1 -2\hat{k} \pm2\hat{k}^{2})
-2\hat{\sigma}(1\mp3\hat{k}+2\hat{k}^{2})
+i\hat{\sigma}^{2}(\mp2+3\hat{k})
\Bigr]+O(n^{-2}).
\label{QNMBSapp}
\end{eqnarray}
%
If we take $\hat{\sigma}=0$, the result reproduces the quasinormal modes of 
the black string obtained in \cite{Emparan:2013moa,Emparan:2015rva}.
The relation of the boosted black string and black ring become clear by observing eq. (\ref{QNMBSapp}). Actually, if we take the large radius
limit of the quasinormal modes $\omega^{(\ell=0)}_{\pm}$ of the black ring in eq. (\ref{QNMBRNLOell0}), we obtain
%
\begin{eqnarray}
&&
\omega^{(\ell=0)}_{\pm} = \hat{m} \pm i\hat{m}(1\mp \hat{m}) \notag \\
&&~~~~~~
+\frac{\hat{m}}{2n}\Bigl[
1\mp6\hat{m}+4\hat{m}^{2} +i(\mp3+\hat{m}\pm2\hat{m}^{2})
\Bigr].
\label{QNMBRNBLOell0RInf}
\end{eqnarray}
%
Comparing eqs. (\ref{QNMBSapp}) and (\ref{QNMBRNBLOell0RInf}) by identifying $\hat{k}=\hat{m}$, we can see that the large radius limit of the black ring corresponds to 
the boosted black string with the boost parameter 
%
\begin{eqnarray}
\sinh{\alpha}&=&\frac{\hat{\sigma}}{\sqrt{n}} \notag \\
&=& \frac{1}{\sqrt{n+1}},
\label{sigmav}
\end{eqnarray}
%
up to $O(1/n)$. This boost parameter corresponds to eq. (\ref{boost}), and this is consistent with the result in \cite{Emparan:2007wm}
\footnote{
In \cite{Emparan:2007wm} they did not use $1/n$ expansion, so the relation (\ref{boost}) obtained in \cite{Emparan:2007wm} should be valid 
at all order in $1/n$ at the large radius limit. 
}. The Gregory-Laflamme mode $\hat{k}_{\text{GL}}$ of the boosted black string, for which the imaginary part of $\omega^{(\ell=0)}_{+}$ vanishes, is obtained as
%
\begin{eqnarray}
\hat{k}_{\text{GL}}=1-\frac{1-\hat{\sigma}^{2}}{2n}+O(n^{-2}).
\end{eqnarray}
%
This Gregory-Laflamme mode of $\hat{\sigma}=1$ reproduces the threshold wave number (\ref{mD}) for the black ring. 

For $\hat{k}=0$ modes, the quasinormal modes up to $1/n$ corrections are
%
\begin{eqnarray}
&&
\omega^{(\hat{k}=0)}_{\pm} = \pm\sqrt{\ell-1}-i(\ell-1) \notag \\
&&~~~~~~
+\frac{1}{2n}\Bigl[
\pm(3\ell-4-\hat{\sigma}^{2})\sqrt{\ell-1}-i(\ell-1)(2\ell-4-\hat{\sigma}^{2})
\Bigr]+O(n^{-2}).
\label{QNMBSapp2}
\end{eqnarray}
%
Using eq. (\ref{sigmav}), the quasinormal mode (\ref{QNMBSapp2}) reproduces eq. (\ref{QNMNLOm0}) at the large radius limit $R=\infty$.

\section{Trivial perturbations in ring coordinate} \label{A}

In this appendix we study trivial perturbations in the $D=n+4$ dimensional ring coordinate given by
%
\begin{eqnarray}
ds^{2}=-dt^{2} +\frac{R^{2}}{(R+r\cos{\theta})^{2}}\Biggl[
\frac{R^{2}dr^{2}}{R^{2}-r^{2}}
 + (R^{2}-r^{2})d\Phi^{2}
+r^{2}(d\theta^{2} +\sin^{2}{\theta}d\Omega^{2}_{n})
\Biggr].
\label{ringmApp}
\end{eqnarray}
%
There are two trivial perturbations, axisymmetric and non-axisymmetric trivial perturbations. 

\paragraph{Axisymmetric trivial perturbation}
The axisymmetric trivial perturbation is the redefinition of the ring radius $R$. The perturbations
%
\begin{eqnarray}
R\rightarrow R + \delta R,
\end{eqnarray}
%
and
%
\begin{eqnarray}
r\rightarrow r+ \delta R\left(\frac{r}{R}+\frac{R^{2}-r^{2}}{R^{2}}\cos{\theta}\right),~~
\theta\rightarrow \theta-\delta R\frac{\sin{\theta}}{r} 
\end{eqnarray}
%
do not change the form of the metric (\ref{ringmApp}) up to $O(\delta R)$. 

As another trivial axisymmetric perturbation, there is a perturbation which becomes trivial at the large $D$ limit. If we consider the perturbation, 
%
\begin{eqnarray}
\Phi\rightarrow \Phi +\delta \Phi
\end{eqnarray}
%
and 
%
\begin{eqnarray}
r\rightarrow r + \delta\Phi\frac{(R^{2}-r^{2})(r+R\cos{\theta})}{R(R+r\cos{\theta})},~~
\theta \rightarrow \theta - \delta\Phi\frac{(R^{2}-r^{2})\sin{\theta}}{r(R+r\cos{\theta})},
\label{rcBR}
\end{eqnarray}
%
the metric does change its form by
%
\begin{eqnarray}
\delta (ds^{2})=-\delta\Phi\frac{2R^{2}}{(R^{2}-r^{2})(R+r\cos{\theta})^{4}}\Bigl(
R^{2}\cos{\theta}dr +r (Rdr-(R^{2}-r^{2})\sin{\theta}d\theta )
\Bigr)^{2}.
\label{metdef}
\end{eqnarray}
%
Thus this transformation is not a trivial perturbation. However, if we consider the large $D$ limit, this transformation becomes trivial.
Taking $\delta\Phi=\hat{\delta\Phi}/n$ and using $\sR=(r/r_{0})^{n}$, the metric deformation (\ref{metdef}) becomes
%
\begin{eqnarray}
\delta (ds^{2}) = \frac{\hat{\delta\Phi}}{n}\frac{2R^{2}(R^{2}-1)}{(R+\cos{\theta})^{4}}d\theta^{2}+O(n^{-2}),
\label{dsdef}
\end{eqnarray}
%
where we set to $r_{0}=1$, and we neglect $dr^{2}$ and $drdz$ terms since $dr\simeq d\sR/(n\sR)$ is higher order in $1/n$. 
The deformation (\ref{dsdef}) can be absorbed into the $1/n$ redefinition of $\theta$.
So the transformation (\ref{rcBR}) becomes trivial at the large $D$ limit. 
We can see that $\sR=(r/r_{0})^{n}$ changes its form by the transformation (\ref{rcBR}) to
%
\begin{eqnarray}
\sR \rightarrow \sR ~\text{exp}\Biggl[
\hat{\delta\Phi}\left( R^{2}-1 -\frac{(R^{2}-1)^{2}}{R(R+\cos{\theta})} \right)
\Biggr].
\end{eqnarray}
%
So, observing the solution eq. (\ref{Psol}), we find that the integration constants $p_{0}$ and $d_{0}$ in eq. (\ref{Psol}) 
represent the transformation (\ref{rcBR}) and the $O(1/n)$ redefinition of $r_{0}$. 
So $p_{0}$ and $d_{0}$ in eq. (\ref{Psol}) do not have physical degree of freedom.

\paragraph{Non-axisymmetric trivial perturbations}
Let us consider the perturbation given by
%
\begin{eqnarray}
r\rightarrow r +\epsilon e^{im_{\Phi}\Phi}\frac{\sqrt{R^{2}-r^{2}}}{R}(r+R\cos{\theta}),~~
z\rightarrow z +\epsilon e^{im_{\Phi}\Phi}\frac{\sqrt{R^{2}-r^{2}}}{r}\sin{\theta}.
\end{eqnarray}
%
For $m_{\Phi}=1$, the metric does not change the form under this non-axisymmetric perturbation up to $O(\epsilon)$. 
So the stationary deformation of the black ring with $m_{\Phi}=1$ is just a trivial deformation. Thus it might be 
reasonable to regard that the perturbations with $m_{\Phi}=1$ and $\ell=0$ does not have physical degree of freedom.

\section{$1/D$ corrections of effective equations} \label{C}

In this appendix we show $1/n$ corrections to the effective equations (\ref{Deq1}), (\ref{Deq2}) and (\ref{Deq3}). 
The effective equations up to $O(1/n)$ are
%
\begin{eqnarray}
\pv p_{v} -\frac{H'(z)}{2\hat{\kappa}H(z)}\pz p_{v} -\frac{\pp^{2} p_{v}}{2\hat{\kappa}G(z)^{2}} 
+\frac{\pp p_{\phi}}{G(z)^{2}}+\frac{H'(z)}{H(z)}p_{z}+\frac{\Delta_{v}}{n}=0,
\label{Deq1app}
\end{eqnarray}
%
%
\begin{eqnarray}
&&
\pv p_{\phi} -\frac{H'(z)}{2\hat{\kappa}H(z)}\pz p_{\phi} -\frac{\pp^{2}p_{\phi}}{2\hat{\kappa}G(z)^{2}} 
+\frac{1}{G(z)^{2}}\pp \Biggl[ \frac{p_{\phi}^{2}}{p_{v}} \Biggr] \notag \\
&&~~~~~
-\frac{4\hat{\kappa}^{2}G(z)H(z)^{2}+2G'(z)H(z)H'(z)}{4\hat{\kappa}^{2}G(z)H(z)^{2}}\pp p_{v}
 \notag \\
&&~~~~~
+\frac{H'(z)}{H(z)}\frac{p_{z}p_{\phi}}{p_{v}}
+\frac{G'(z)H'(z)}{\hat{\kappa}G(z)H(z)}p_{\phi}+\frac{\Delta_{\phi}}{n}=0,
\label{Deq2app}
\end{eqnarray}
%
and
%
\begin{eqnarray}
&&
\pv p_{z} -\frac{H'(z)}{2\hat{\kappa}H(z)}\pz p_{z} -\frac{\pp^{2}p_{z}}{2\hat{\kappa}G(z)^{2}}
+\pz p_{v} +\frac{1}{G(z)^{2}}\pp \Biggl[ \frac{p_{\phi}p_{z}}{p_{v}} \Biggr] \notag \\
&&~~~~~~~
+\frac{H'(z)}{H(z)}\frac{p^{2}_{z}}{p_{v}}
-\frac{G'(z)}{G(z)^{3}}\frac{p^{2}_{\phi}}{p_{v}} +\frac{G'(z)}{\hat{\kappa}G(z)^{3}}\pp p_{\phi} \notag \\
&&~~~~~~~
+\frac{H(z)G'(z)^{2}H'(z)+G(z)(G'(z)-H(z)H'(z)G''(z))}{4\hat{\kappa}^{2}G(z)^{2}H(z)^{2}}p_{v} \notag \\
&&~~~~~~~
-\frac{1-2H'(z)^{2}}{2\hat{\kappa}H(z)^{2}}p_{z}+\frac{\Delta_{z}}{n}=0.
\label{Deq3app}
\end{eqnarray}
%
The corrections terms $\Delta_{v}$, $\Delta_{\phi}$ and $\Delta_{z}$ are given by
%
\begin{eqnarray}
&&
\Delta_{v}=
-\frac{\pp^{2} p_{v} \left(5 G' H'+4 \hat{\kappa} ^2 G H\right)}{8 \hat{\kappa} ^3 G^3 H}
-\frac{\pz p_{v} \left(G'-12 \hat{\kappa} ^2 G H H'\right)}{8 \hat{\kappa} ^3 GH^2} \\ \notag
&& 
+\frac{1}{p_{v}} \Biggl(
-\frac{G' H' p_{\phi}^2}{\hat{\kappa}  G^3 H}+p_{\phi} \left(\frac{H' \pz p_{\phi}}{2 \hat{\kappa}  G^2 H}+\frac{H'
   \pp p_{z}}{\hat{\kappa}  G^2 H}+\frac{3 \pp^{2} p_{\phi}}{2 \hat{\kappa}  G^4}\right)
+\frac{H' \pp p_{\phi} p_{z}}{\hat{\kappa}  G^2 H} \notag \\
&&~~~~~~~~~~
+\frac{3
   (\pp p_{\phi})^2}{2 \hat{\kappa}  G^4}+\left(\frac{1}{2 \hat{\kappa}  H^2}-2 \hat{\kappa} \right) p_{z}^2
\Biggr) 
+\frac{\pp p_{\phi} \left(\frac{2 G' H'}{\hat{\kappa} ^2
   H}-G\right)}{G^3} \notag \\
&&
+p_{z} \left(\frac{G'}{G}+\frac{\left(\frac{1}{\hat{\kappa} ^2}-10 H^2\right) H'}{2 H^3}\right) 
-\frac{H' \pz\pp p_{\phi}}{2 \hat{\kappa} ^2 G^2 H} 
-\frac{H' \pp^{2} p_{z}}{2 \hat{\kappa} ^2 G^2 H} 
+\frac{3 (\pp p_{v})^2 p_{\phi}^2}{2 \hat{\kappa}  G^4 p_{v}^3}
 \notag \\
&&
+\frac{-\frac{H' \pp p_{v} p_{\phi} p_{z}}{\hat{\kappa}  G^2 H} 
+p_{\phi}^2
   \left(-\frac{H' \pz p_{v}}{4 \hat{\kappa}  G^2 H}-\frac{3 \pp^{2} p_{v}}{4 \hat{\kappa}  G^4}\right) 
-\frac{3 \pp p_{v} \pp p_{\phi} p_{\phi}}{\hat{\kappa} 
   G^4}}{p_{v}^2} 
   +\left(3-\frac{1}{2 \hat{\kappa} ^2 H^2}\right)
   \pz p_{z} \notag \\
&&
-\frac{\hat{\kappa}  G^4
   \pz^{2} p_{v}+\pp^{3} p_{\phi}}{2 \hat{\kappa} ^2 G^4}
   +\frac{H' p_{v} \left(H (G')^{2} H'+G
   \left(G'-H G'' H'\right)\right)}{4 \hat{\kappa} ^3 G^2 H^3},
\end{eqnarray}
%
%
\begin{eqnarray}
&&
\Delta_{\phi}=
 \frac{G''}{\hat{\kappa}  G}p_{\phi}
 -\frac{4 G H G' H' \hat{\kappa} ^2+8 G H^2 G'' \hat{\kappa} ^2-3 \left(4 \hat{\kappa} ^2 H^2-1\right) (G')^{2}}{16 \hat{\kappa} ^4 G^2 H^2}\pp p_{v} 
 \notag \\
&&
-\frac{ 12 \hat{\kappa} ^2 \left((G')^{2}-G G''\right) H^3+8 \hat{\kappa} ^2 G G' H' H^2-3 \left((G')^{2}-G G''\right)
   H-3 G G' H'}{8 \hat{\kappa} ^3 G^2 H^3}p_{\phi}\log{p_{v}}
\notag \\
&&
+\frac{3 p_{\phi}^3\log{p_{v}} (\pp p_{v})^2}{\hat{\kappa}  G^4p_{v}^4}
-\frac{p_{\phi}^3(\pp p_{v})^2}{\hat{\kappa}  G^4 p_{v}^4}
+\frac{3  p_{\phi}^2\log{p_{v}} (\pp p_{v})^3}{2 \hat{\kappa} ^2 G^4 p_{v}^4}+\frac{ p_{\phi}^2(\pp p_{v})^3}{4 \hat{\kappa} ^2 G^4 p_{v}^4}
+\frac{\left(\frac{H'}{H}-\frac{2 G'}{G}\right) \pp p_{z}}{2 \hat{\kappa} }
 \notag \\
&&
-\frac{\left(2 G H \hat{\kappa} ^2+G' H'\right)
   \pp^{2} p_{\phi}}{4 \hat{\kappa} ^3 G^3 H}
+\frac{G' \pz p_{\phi}}{2 \hat{\kappa}  G}  
-\frac{\pz^{2} p_{\phi}}{2 \hat{\kappa} }
-\frac{3 G' H' \pv p_{\phi}}{4 \hat{\kappa} ^2 G H}
\notag \\
&&
+\log{p_{v}} \Biggl(
\frac{\left(4 \hat{\kappa} ^2 \left((G')^{2}-G G''\right)
   H^3+\left(G G''-(G')^{2}\right) H-G G' H'\right) \pp p_{v}}{8 \hat{\kappa} ^4 G^2 H^3} \notag \\
&&~~~~
+\frac{\left(\left(4 \hat{\kappa} ^2 H^2-1\right) G'-2 \hat{\kappa} ^2 G H
   H'\right) \pp p_{z}}{4 \hat{\kappa} ^3 G H^2}-\frac{\left(2 G H \hat{\kappa} ^2+G' H'\right) \pp^{2} p_{\phi}}{2 \hat{\kappa} ^3 G^3 H}+\frac{H'
   \pz p_{\phi}}{2 \hat{\kappa}  H}\Biggr) \notag \\
&&
+\frac{\Delta_{\phi}^{(3)}}{p_{v}^3}
+\frac{\Delta_{\phi}^{(2)}}{p_{v}^2}
+\frac{\Delta_{\phi}^{(1)}}{p_{v}},
\end{eqnarray}
%
and
%
\begin{eqnarray}
&&
\Delta_{z}=
\left(2-\frac{G'
   H'}{4 \hat{\kappa} ^2 G H}\right) \pv p_{z}
   -\frac{\pz^{2} p_{z}}{2 \hat{\kappa} }
+p_{z} \Biggl(
\frac{-G^2+H^2 G'' G+H^2 (G')^{2}}{2 \hat{\kappa}  G^2 H^2} \notag \\
&&~~~~~~
+\frac{\log{p_{v}} \left(-4 \hat{\kappa} ^2 H (H')^{2} G^2+\left(G'
   H'-H(H')^{2}  G''\right) G+H (H')^{2} (G')^{2}\right)}{4 \hat{\kappa} ^3 G^2 H^3}\Biggr) \notag \\
&&   
+p_{v} \Biggl(
\frac{\log{p_{v}} G'
   \left(-4 \hat{\kappa} ^2 \left((G')^{2}-G G''\right) H^3+\left((G')^{2}-G G''\right) H+G G' H'\right)}{8 \hat{\kappa} ^4 G^3 H^3} \notag \\
&&
+\frac{-2 \hat{\kappa} ^2 \left(2 (G')^{3}-3 G G''
   G'+G^2 G^{(3)}(z)\right) H^3+\left((G')^{3}-G G' G''\right) H+G (G')^{2} H'}{8 \hat{\kappa} ^4 G^3 H^3}\Biggr) \notag \\
&&
+\frac{\left(3 H H' (G')^{2}+G \left(\left(4 \hat{\kappa} ^2
   H^2+1\right) G'-H H' G''\right)\right) \pp p_{\phi}}{4 \hat{\kappa} ^3 G^4 H^2} 
+\frac{\left(G H'-H G'\right) \pz p_{z}}{2 \hat{\kappa}  G H} \notag \\
&& 
+\frac{\left(4 G H G' H' \hat{\kappa} ^2-8 G H^2 G'' \hat{\kappa} ^2+\left(4 \hat{\kappa}
   ^2 H^2-1\right) (G')^{2}\right) \pz p_{v}}{16 \hat{\kappa} ^4 G^2 H^2} 
-\frac{\left(2 G H \hat{\kappa} ^2+G' H'\right)
   \pz\pp p_{\phi}}{4 \hat{\kappa} ^3 G^3 H} \notag \\
&&   
+\log{p_{v}} \Biggl(
\frac{\left(7 H H' (G')^{2}+3 G \left(G'-H H' G''\right)\right) \pp p_{\phi}}{8 \hat{\kappa} ^3
   G^4 H^2}-\frac{\left(2 G H \hat{\kappa} ^2+G' H'\right) \pp^{2} p_{z}}{4 \hat{\kappa} ^3 G^3 H} \notag \\
&&
+\frac{H' \pz p_{z}}{\hat{\kappa} 
   H}+\frac{\pz\pp p_{\phi}}{2 \hat{\kappa}  G^2}
\Biggr) 
+\frac{\Delta_{z}^{(4)}}{p_{v}^4} 
+\frac{\Delta_{z}^{(3)}}{p_{v}^3} 
+\frac{\Delta_{z}^{(2)}}{p_{v}^2} 
+\frac{\Delta_{z}^{(1)}}{p_{v}}.
\end{eqnarray}
%
The coefficients $\Delta_{\phi}^{(1,2,3)}$ and $\Delta_{z}^{(1,2,3,4)}$ have messy forms as
%
\begin{eqnarray}
&&
\Delta_{\phi}^{(3)}=
\left(-\frac{\log{p_{v}} \pp^{2} p_{v}}{\hat{\kappa}  G^4}-\frac{H' \pz p_{v}}{2
   \hat{\kappa}  G^2 H}\right) p_{\phi}^3 
+\Biggl(\pp p_{v} \left(\frac{3 \pp p_{\phi}}{2 \hat{\kappa}  G^4}-\frac{\pp^{2} p_{v}}{2 \hat{\kappa} ^2 G^4}\right)-\frac{2
   \pp p_{v} \pv p_{v}}{\hat{\kappa}  G^2} \notag \\
&&~~~~~~~
+\log{p_{v}} \left(\pp p_{v} \left(-\frac{6 \pp p_{\phi}}{\hat{\kappa}  G^4}-\frac{3 \pp^{2} p_{v}}{2 \hat{\kappa} ^2
   G^4}\right)-\frac{H' \pp p_{v} \pz p_{v}}{2 \hat{\kappa} ^2 G^2 H}+\frac{\pp p_{v} \pv p_{v}}{\hat{\kappa}  G^2}\right)\Biggr) p_{\phi}^2
   \notag \\
&&~~~~~   
+p_{z}
   \left(p_{\phi}^2 \left(\frac{2 H' \pp p_{v}}{\hat{\kappa}  G^2 H}-\frac{2 \log{p_{v}} H' \pp p_{v}}{\hat{\kappa}  G^2 H}\right)-\frac{\log{p_{v}}
   p_{\phi} H' (\pp p_{v})^2}{2 \hat{\kappa} ^2 G^2 H}\right) \notag \\
&&~~~~~
+\left(-\frac{3
   \log{p_{v}} \pp p_{\phi} (\pp p_{v})^2}{\hat{\kappa} ^2 G^4}-\frac{\pp p_{\phi} (\pp p_{v})^2}{2 \hat{\kappa} ^2 G^4}\right) p_{\phi},   
\end{eqnarray}
%
%
\begin{eqnarray}
&&
\Delta_{\phi}^{(2)}= 
\left(-\frac{\log{p_{v}}
   G' H'}{2 \hat{\kappa}  G^3 H}-\frac{G' H'}{2 \hat{\kappa}  G^3 H}\right) p_{\phi}^3
+\Biggl(\frac{\left(\frac{3 G' H'}{\hat{\kappa} ^2 H}-4 G\right) \pp p_{v}}{4
   G^3}-\frac{3 H' \pp p_{z}}{4 \hat{\kappa}  G^2 H}+\frac{\pp^{3} p_{v}}{8 \hat{\kappa} ^2 G^4} \notag \\
&&
+\frac{3 H' \pz p_{\phi}}{4 \hat{\kappa}  G^2 H}+\frac{H'
   \pz\pp p_{v}}{8 \hat{\kappa} ^2 G^2 H}+\frac{3 \pv\pp p_{v}}{4 \hat{\kappa}  G^2}+\log{p_{v}} \Biggl(-\frac{3 G' H' \pp p_{v}}{2 \hat{\kappa} ^2 G^3 H}+\frac{H'
   \pp p_{z}}{\hat{\kappa}  G^2 H}+\frac{3 \pp^{2} p_{\phi}}{2 \hat{\kappa}  G^4} \notag \\
&&
+\frac{\pp^{3} p_{v}}{4 \hat{\kappa} ^2 G^4}+\frac{H' \pz\pp p_{v}}{4 \hat{\kappa} ^2 G^2
   H}-\frac{\pv\pp p_{v}}{2 \hat{\kappa}  G^2}\Biggr)\Biggr) p_{\phi}^2
+\Biggl(2 \hat{\kappa}+\left(\frac{1}{2 \hat{\kappa}  H^2}-2 \hat{\kappa} \right) \log{p_{v}}-\frac{1}{2 H^2 \hat{\kappa} }\Biggr)
   p_{z}^2 p_{\phi} \notag \\
&&
+\Biggl(\frac{3 \pp p_{\phi} \pp^{2} p_{v}}{4 \hat{\kappa} ^2 G^4}+\frac{\pp p_{v} \pp^{2} p_{\phi}}{4 \hat{\kappa} ^2
   G^4}+\frac{H' \pp p_{\phi} \pz p_{v}}{4 \hat{\kappa} ^2 G^2 H}-\frac{H' \pp p_{v} \pz p_{\phi}}{4 \hat{\kappa} ^2 G^2 H}+\frac{3
   \pp p_{\phi} \pv p_{v}}{2 \hat{\kappa}  G^2}+\frac{5 \pp p_{v} \pv p_{\phi}}{2 \hat{\kappa}  G^2} \notag \\
&&
+\log{p_{v}} \Biggl(\frac{3
   (\pp p_{\phi})^2}{\hat{\kappa}  G^4}+\frac{3 \pp^{2} p_{v} \pp p_{\phi}}{2 \hat{\kappa} ^2 G^4}+\frac{H' \pz p_{v} \pp p_{\phi}}{2 \hat{\kappa} ^2 G^2
   H} \left(\frac{H' \pp p_{z}}{2 \hat{\kappa} ^2 G^2 H}+\frac{3
   \pp^{2} p_{\phi}}{2 \hat{\kappa} ^2 G^4}\right) \notag \\
&& 
-\frac{\pv p_{v} \pp p_{\phi}}{\hat{\kappa}  G^2}+\pp p_{v}
+\frac{H' \pp p_{v} \pz p_{\phi}}{2 \hat{\kappa} ^2 G^2 H}-\frac{\pp p_{v} \pv p_{\phi}}{\hat{\kappa} 
   G^2}\Biggr)\Biggr) p_{\phi}+\frac{3 \log{p_{v}} \pp p_{v} (\pp p_{\phi})^2}{2 \hat{\kappa} ^2 G^4}
\notag \\
&&
+\frac{\pp p_{v} (\pp p_{\phi})^2}{4 \hat{\kappa} ^2
   G^4} 
+p_{z} \Biggl(\frac{\log{p_{v}} H' \pp p_{v} \pp p_{\phi}}{2 \hat{\kappa} ^2 G^2 H}+p_{\phi} \Biggl(-\frac{3 H' \pp p_{\phi}}{2
   \hat{\kappa}  G^2 H}-\pz p_{v}+\frac{H' \pv p_{v}}{\hat{\kappa}  H} \notag \\
&&
+\log{p_{v}} \left(\frac{2 H' \pp p_{\phi}}{\hat{\kappa}  G^2 H}+\frac{H'
   \pp^{2} p_{v}}{4 \hat{\kappa} ^2 G^2 H}+\left(\frac{1}{4 \hat{\kappa} ^2 H^2}-1\right) \pz p_{v}-\frac{H' \pv p_{v}}{2 \hat{\kappa} 
   H}\right)\Biggr)\Biggr), 
\end{eqnarray}
%
%
\begin{eqnarray}
&&
\Delta_{\phi}^{(1)}=
-\frac{\left(2 G H \hat{\kappa} ^2+G' H'\right) \pp p_{v} \pp p_{\phi}}{2 \hat{\kappa} ^3 G^3 H}
-\frac{3\pp^{2} p_{\phi} \pp p_{\phi}}{4 \hat{\kappa} ^2 G^4}
-\frac{H' \pz p_{\phi} \pp p_{\phi}}{4 \hat{\kappa} ^2 G^2 H}
-\frac{3\pv p_{\phi} \pp p_{\phi}}{2 \hat{\kappa}  G^2} \notag \\
&&
+\log{p_{v}} \left(\pp p_{\phi} \left(-\frac{H' \pp p_{z}}{2 \hat{\kappa} ^2 G^2
   H}-\frac{3 \pp^{2} p_{\phi}}{2 \hat{\kappa} ^2 G^4}\right)-\frac{H' \pp p_{\phi} \pz p_{\phi}}{2 \hat{\kappa} ^2 G^2 H}+\frac{\pp p_{\phi}
   \pv p_{\phi}}{\hat{\kappa}  G^2}\right) \notag \\
&&
+p_{z} \Biggl(p_{\phi} \left(\frac{\left(16 \hat{\kappa} ^2 H^2-3\right) G'}{4 \hat{\kappa} ^2 G H^2}+\frac{\log{p_{v}} \left(-8
   \hat{\kappa} ^2 G' H^3-8 \hat{\kappa} ^2 G H' H^2+2 G' H+G H'\right)}{4 \hat{\kappa} ^2 G H^3}\right) \notag \\
&&
+\pz p_{\phi}-\frac{H' \pv p_{\phi}}{\hat{\kappa} 
   H}+\log{p_{v}} \left(-\frac{H' \pp^{2} p_{\phi}}{4 \hat{\kappa} ^2 G^2 H}+\left(1-\frac{1}{4 \hat{\kappa} ^2 H^2}\right) \pz p_{\phi}+\frac{H'
   \pv p_{\phi}}{2 \hat{\kappa}  H}\right)\Biggr) \notag \\
&&
+p_{\phi} \Biggl(\frac{\left(4 G+\frac{G' H'}{\hat{\kappa} ^2 H}\right) \pp p_{\phi}}{2
   G^3}-\frac{\pp^{3} p_{\phi}}{4 \hat{\kappa} ^2 G^4}+\pz p_{z}-\frac{H' \pz\pp p_{\phi}}{4 \hat{\kappa} ^2 G^2 H}-\frac{H' \pv p_{z}}{\hat{\kappa}
    H}-\frac{3 \pv\pp p_{\phi}}{2 \hat{\kappa}  G^2} \notag \\
&&
+\log{p_{v}} \Biggl(\frac{3 G' H' \pp p_{\phi}}{\hat{\kappa} ^2 G^3 H}-\frac{H' \pp^{2} p_{z}}{4
   \hat{\kappa} ^2 G^2 H}-\frac{\pp^{3} p_{\phi}}{2 \hat{\kappa} ^2 G^4}+\left(1-\frac{1}{4 \hat{\kappa} ^2 H^2}\right) \pz p_{z} 
\notag \\
&&~~~~~~~~~~~~~~~
-\frac{H' \pz\pp p_{\phi}}{2 \hat{\kappa} ^2
   G^2 H}+\frac{H' \pv p_{z}}{2 \hat{\kappa}  H}+\frac{\pv\pp p_{\phi}}{\hat{\kappa}  G^2}\Biggr)\Biggr),
\end{eqnarray}
%
%
\begin{eqnarray}
&&
\Delta_{z}^{(4)}=
\frac{3 \pp p_{v} \pz p_{v} p_{\phi}^3}{2 \hat{\kappa}  G^4}+\frac{3 \pp p_{v}^2
   \pz p_{v} p_{\phi}^2}{4 \hat{\kappa} ^2 G^4} \notag \\
&&
+p_{z} \left(\left(\frac{3 \log{p_{v}} (\pp p_{v})^2}{\hat{\kappa}  G^4}-\frac{5 \pp p_{v}^2}{2 \hat{\kappa} 
   G^4}\right) p_{\phi}^2+\left(\frac{3 \log{p_{v}} (\pp p_{v})^3}{2 \hat{\kappa} ^2 G^4}-\frac{(\pp p_{v})^3}{2 \hat{\kappa} ^2 G^4}\right)
   p_{\phi}\right),
\end{eqnarray}
%
%
\begin{eqnarray}
&&
\Delta_{z}^{(3)}=
\left(\frac{2 \log{p_{v}} G' \pp p_{v}}{\hat{\kappa}  G^5}+\frac{G'
   \pp p_{v}}{2 \hat{\kappa}  G^5}-\frac{\pz\pp p_{v}}{2 \hat{\kappa}  G^4}\right) p_{\phi}^3+\Biggl(\frac{G' \pp p_{v}^2}{2 \hat{\kappa} ^2 G^5}-\frac{3
   \pz p_{\phi} \pp p_{v}}{2 \hat{\kappa}  G^4} \notag \\
&&
+\left(\frac{3 \pp p_{z}}{2 \hat{\kappa}  G^4}-\frac{\pz\pp p_{v}}{2 \hat{\kappa} ^2 G^4}\right)
   \pp p_{v}-\frac{H' \pz p_{v}^2}{4 \hat{\kappa} ^2 G^2 H}+\log{p_{v}} \left(\frac{3 G' (\pp p_{v})^2}{2 \hat{\kappa} ^2 G^5}-\frac{2 \pp p_{v}
   \pp p_{z}}{\hat{\kappa}  G^4}\right) \notag \\
&&
+\left(-\frac{3 \pp p_{\phi}}{2 \hat{\kappa}  G^4}-\frac{\pp^{2} p_{v}}{4 \hat{\kappa} ^2 G^4}\right)
   \pz p_{v}+\frac{\pz p_{v} \pv p_{v}}{2 \hat{\kappa}  G^2}\Biggr) p_{\phi}^2+\Biggl(-\frac{3 \log{p_{v}} \pp p_{z} (\pp p_{v})^2}{2 \hat{\kappa} ^2
   G^4}+\frac{\pp p_{z} (\pp p_{v})^2}{2 \hat{\kappa} ^2 G^4} \notag \\
&&
-\frac{\pz p_{\phi} \pp p_{v}^2}{2 \hat{\kappa} ^2 G^4}-\frac{\pp p_{\phi}
   \pz p_{v} \pp p_{v}}{\hat{\kappa} ^2 G^4}\Biggr) p_{\phi} \notag \\
&&
+p_{z}^2 \left(p_{\phi} \left(\frac{5 H' \pp p_{v}}{2 \hat{\kappa}  G^2 H}-\frac{2 \log{p_{v}} H' \pp p_{v}}{\hat{\kappa}  G^2 H}\right)-\frac{\log{p_{v}} H' (\pp p_{v})^2}{2 \hat{\kappa} ^2 G^2 H}\right) \notag \\
&&
+p_{z} \Biggl(\left(-\frac{\log{p_{v}}
   \pp^{2} p_{v}}{\hat{\kappa}  G^4}+\frac{\pp^{2} p_{v}}{2 \hat{\kappa}  G^4}-\frac{H' \pz p_{v}}{\hat{\kappa}  G^2 H}\right) p_{\phi}^2+\Biggl(\pp p_{v} \left(\frac{3
   \pp p_{\phi}}{\hat{\kappa}  G^4}+\frac{\pp^{2} p_{v}}{4 \hat{\kappa} ^2 G^4}\right)+\frac{H' \pp p_{v} \pz p_{v}}{4 \hat{\kappa} ^2 G^2 H}
\notag \\
&&
-\frac{5 \pp p_{v}
   \pv p_{v}}{2 \hat{\kappa}  G^2}+\log{p_{v}} \left(\pp p_{v} \left(-\frac{4 \pp p_{\phi}}{\hat{\kappa}  G^4}-\frac{3 \pp^{2} p_{v}}{2 \hat{\kappa} ^2
   G^4}\right)-\frac{H' \pp p_{v} \pz p_{v}}{2 \hat{\kappa} ^2 G^2 H}+\frac{\pp p_{v} \pv p_{v}}{\hat{\kappa}  G^2}\right)\Biggr) p_{\phi}
\notag \\
&&
-\frac{3 \log
   (p_{v}) (\pp p_{v})^2 \pp p_{\phi}}{2 \hat{\kappa} ^2 G^4}+\frac{\pp p_{v}^2 \pp p_{\phi}}{2 \hat{\kappa} ^2 G^4}\Biggr),
\end{eqnarray}
%
%
\begin{eqnarray}
&&
\Delta_{z}^{(2)}=
\left(2 \hat{\kappa} +\left(\frac{1}{2 \hat{\kappa}  H^2}-2 \hat{\kappa} \right) \log{p_{v}}-\frac{1}{2 H^2 \hat{\kappa} }\right)
   p_{z}^3+\Biggl(-\frac{H' \pp p_{\phi}}{\hat{\kappa}  G^2 H}-\pz p_{v}+\frac{H' \pv p_{v}}{\hat{\kappa}  H} \notag \\
&&
+\log{p_{v}} \left(\frac{H'
   \pp p_{\phi}}{\hat{\kappa}  G^2 H}+\frac{H' \pp^{2} p_{v}}{4 \hat{\kappa} ^2 G^2 H}+\left(\frac{1}{4 \hat{\kappa} ^2 H^2}-1\right) \pz p_{v}-\frac{H'
   \pv p_{v}}{2 \hat{\kappa}  H}\right)\Biggr) p_{z}^2 \notag \\
&&
+\Biggl(\left(\frac{G \left(4 \hat{\kappa} ^2 H^2-1\right)+H G' H'}{2 \hat{\kappa}  G^3 H^2}-\frac{3 \log{p_{v}} G'
   H'}{2 \hat{\kappa}  G^3 H}\right) p_{\phi}^2 
+\frac{3 \pp p_{v} \pv p_{\phi}}{2 \hat{\kappa}  G^2} \notag \\
&&
+\Biggl(\frac{\left(-16 \hat{\kappa} ^2 G H^2+3 G' H' H+G\right) \pp p_{v}}{4 \hat{\kappa} ^2 G^3 H^2}-\frac{2 H'
   \pp p_{z}}{\hat{\kappa}  G^2 H}-\frac{\pp^{2} p_{\phi}}{2 \hat{\kappa}  G^4}+\frac{H' \pz p_{\phi}}{\hat{\kappa}  G^2 H}+\frac{\pv\pp p_{v}}{\hat{\kappa} 
   G^2} \notag \\
&&
+\log{p_{v}} \Biggl(-\frac{\left(-8 \hat{\kappa} ^2 G H^2+6 G' H' H+G\right) \pp p_{v}}{4 \hat{\kappa} ^2 G^3 H^2}+\frac{2 H' \pp p_{z}}{\hat{\kappa} 
   G^2 H}+\frac{\pp^{2} p_{\phi}}{\hat{\kappa}  G^4}+\frac{\pp^{3} p_{v}}{4 \hat{\kappa} ^2 G^4} \notag \\
&&
+\frac{H' \pz\pp p_{v}}{4 \hat{\kappa} ^2 G^2 H} -\frac{\pv\pp p_{v}}{2 \hat{\kappa}
    G^2}\Biggr)\Biggr) p_{\phi}-\frac{(\pp p_{\phi})^2}{2 \hat{\kappa}  G^4}-\frac{\pp p_{v} \pp^{2} p_{\phi}}{4 \hat{\kappa} ^2 G^4}-\frac{H' \pp p_{v}
   \pz p_{\phi}}{4 \hat{\kappa} ^2 G^2 H}+\frac{\pp p_{\phi} \pv p_{v}}{\hat{\kappa}  G^2} \notag \\
&&
+\log{p_{v}} \Biggl(\frac{(\pp p_{\phi})^2}{\hat{\kappa}  G^4}+\frac{3 \pp^{2} p_{v} \pp p_{\phi}}{4 \hat{\kappa} ^2 G^4}+\frac{H' \pz p_{v}
   \pp p_{\phi}}{4 \hat{\kappa} ^2 G^2 H}-\frac{\pv p_{v} \pp p_{\phi}}{2 \hat{\kappa}  G^2}+\pp p_{v} \left(\frac{H' \pp p_{z}}{\hat{\kappa} ^2
   G^2 H}+\frac{3 \pp^{2} p_{\phi}}{4 \hat{\kappa} ^2 G^4}\right) \notag \\
&&
+\frac{H' \pp p_{v} \pz p_{\phi}}{4 \hat{\kappa} ^2 G^2 H}-\frac{\pp p_{v}
   \pv p_{\phi}}{2 \hat{\kappa}  G^2}\Biggr)\Biggr) p_{z}+\frac{3 \log{p_{v}} \pp p_{v} \pp p_{\phi} \pp p_{z}}{2 \hat{\kappa} ^2
   G^4}-\frac{\pp p_{v} \pp p_{\phi} \pp p_{z}}{2 \hat{\kappa} ^2 G^4}+\frac{(\pp p_{\phi})^2 \pz p_{v}}{4 \hat{\kappa} ^2
   G^4} \notag \\
&&
+\frac{\pp p_{v} \pp p_{\phi} \pz p_{\phi}}{2 \hat{\kappa} ^2 G^4}+p_{\phi} \Biggl(\left(\frac{3 \pp p_{\phi}}{2 \hat{\kappa} 
   G^4}+\frac{\pp^{2} p_{v}}{4 \hat{\kappa} ^2 G^4}\right) \pz p_{\phi}+\pz p_{v} \left(\frac{\pp^{2} p_{\phi}}{4 \hat{\kappa} ^2 G^4}+\frac{H'
   \pz p_{\phi}}{2 \hat{\kappa} ^2 G^2 H}\right) \notag \\
&&
-\frac{H' \pp p_{v} \pz p_{z}}{4 \hat{\kappa} ^2 G^2 H}
+\pp p_{\phi}
   \left(\frac{\pz\pp p_{v}}{2 \hat{\kappa} ^2 G^4}-\frac{\pp p_{z}}{\hat{\kappa}  G^4}\right)+\pp p_{v} \left(-\frac{G' \pp p_{\phi}}{\hat{\kappa} ^2
   G^5}-\frac{\pp^{2} p_{z}}{4 \hat{\kappa} ^2 G^4}+\frac{\pz\pp p_{\phi}}{2 \hat{\kappa} ^2 G^4}\right) \notag \\
&&
+\left(\frac{\pp p_{z}}{\hat{\kappa} 
   G^2}-\frac{\pz p_{\phi}}{2 \hat{\kappa}  G^2}\right) \pv p_{v}
-\frac{\pz p_{v} \pv p_{\phi}}{2 \hat{\kappa}  G^2}
+\frac{3 \pp p_{v}
   \pv p_{z}}{2 \hat{\kappa}  G^2}+\log{p_{v}} \Biggl(\frac{2 \pp p_{\phi} \pp p_{z}}{\hat{\kappa}  G^4}+\frac{3 \pp^{2} p_{v}
   \pp p_{z}}{4 \hat{\kappa} ^2 G^4} \notag \\
&&
+\frac{H' \pz p_{v} \pp p_{z}}{4 \hat{\kappa} ^2 G^2 H}-\frac{\pv p_{v} \pp p_{z}}{2 \hat{\kappa} 
   G^2}
+\pp p_{v} \left(\frac{3 \pp^{2} p_{z}}{4 \hat{\kappa} ^2 G^4}-\frac{3 G' \pp p_{\phi}}{\hat{\kappa} ^2 G^5}\right)+\frac{H' \pp p_{v}
   \pz p_{z}}{4 \hat{\kappa} ^2 G^2 H}-\frac{\pp p_{v} \pv p_{z}}{2 \hat{\kappa}  G^2}\Biggr)\Biggr) \notag \\
&&
+p_{\phi}^2 \Biggl(-\frac{3 G'
   \pp p_{\phi}}{2 \hat{\kappa}  G^5}-\frac{G' \pp^{2} p_{v}}{2 \hat{\kappa} ^2 G^5}
-\frac{\pp^{2} p_{z}}{4 \hat{\kappa}  G^4}+\frac{\left(G \left(H'^4+6 \hat{\kappa} ^2
   H^2-1\right)-12 \hat{\kappa} ^2 H^3 G' H'\right) \pz p_{v}}{16 \hat{\kappa} ^4 G^3 H^4}
\notag \\
&&
+\frac{H' \pz p_{z}}{2 \hat{\kappa}  G^2 H} 
+\frac{3
   \pz\pp p_{\phi}}{4 \hat{\kappa}  G^4}   
+\frac{H' \pz^{2} p_{v} G^2+H m^{(0,1,2)}(u,z,x)}{8 \hat{\kappa} ^2 G^4 H}-\frac{G' \pv p_{v}}{2 \hat{\kappa}  G^3} \notag \\
&&
+\log{p_{v}}
   \Biggl(-\frac{3 G' \pp p_{\phi}}{\hat{\kappa}  G^5}-\frac{3 G' \pp^{2} p_{v}}{4 \hat{\kappa} ^2 G^5}+\frac{\pp^{2} p_{z}}{2 \hat{\kappa}  G^4}
-\frac{G' H'
   \pz p_{v}}{4 \hat{\kappa} ^2 G^3 H}+\frac{G' \pv p_{v}}{2 \hat{\kappa}  G^3}\Biggr)-\frac{\pv\pz p_{v}}{4 \hat{\kappa}  G^2}\Biggr),
\end{eqnarray}
%
%
\begin{eqnarray}
&&
\Delta_{z}^{(1)}=
\Biggl(\frac{4 G
   \left(\left(2 \hat{\kappa} ^2 H^2-1\right) G'+H H' G''\right)-5 H (G')^{2} H'}{4 \hat{\kappa} ^2 G^4 H^2} \notag \\
&&
+\frac{\log{p_{v}} \left(G \left(\left(4 \hat{\kappa} ^2 H^2-1\right)
   G'+H H' G''\right)-3 H (G')^{2} H'\right)}{2 \hat{\kappa} ^2 G^4 H^2}\Biggr) p_{\phi}^2 \notag \\
&&
+\Biggl(\frac{\left(3 H H' (G')^{2}+G \left(\left(3-8 \hat{\kappa} ^2 H^2\right)
   G'-3 H H' G''\right)\right) \pp p_{v}}{8 \hat{\kappa} ^3 G^4 H^2}+\frac{\left(8 G-\frac{G' H'}{\hat{\kappa} ^2 H}\right) \pp p_{z}}{4 G^3}
\notag \\
&&
+\frac{G'
   \pp^{2} p_{\phi}}{\hat{\kappa} ^2 G^5}+\frac{\left(-8 \hat{\kappa} ^2 G H^2+6 G' H' H+G\right) \pz p_{\phi}}{4 \hat{\kappa} ^2 G^3 H^2}-\frac{H'
   \pz^{2} p_{\phi} G^2+H \pz\pp^{2}p_{\phi}}{4 \hat{\kappa} ^2 G^4 H}+\frac{G' \pv p_{\phi}}{\hat{\kappa}  G^3} \notag \\
&&
-\frac{\pv\pp p_{z}}{\hat{\kappa} 
   G^2}+\log{p_{v}} \Biggl(\frac{\left(-8 \hat{\kappa} ^2 G H^2+6 G' H' H+G\right) \pp p_{z}}{4 \hat{\kappa} ^2 G^3 H^2}+\frac{3 G' \pp^{2} p_{\phi}}{2
   \hat{\kappa} ^2 G^5}-\frac{\pp^{3} p_{z}}{4 \hat{\kappa} ^2 G^4}+\frac{G' H' \pz p_{\phi}}{2 \hat{\kappa} ^2 G^3 H} \notag \\
&&
-\frac{H' \pz\pp p_{z}}{4 \hat{\kappa} ^2
   G^2 H}-\frac{G' \pv p_{\phi}}{\hat{\kappa}  G^3}+\frac{\pv\pp p_{z}}{2 \hat{\kappa}  G^2}\Biggr)+\frac{\pv\pz p_{\phi}}{2 \hat{\kappa}  G^2}\Biggr)
   p_{\phi}+\frac{G' (\pp p_{\phi})^2}{\hat{\kappa} ^2 G^5}-\frac{H' (\pz p_{\phi})^2}{4 \hat{\kappa} ^2 G^2 H} \notag \\
&&
+p_{z}^2 \left(\frac{\left(8-\frac{1}{\hat{\kappa}
   ^2 H^2}\right) G'}{4 G}-\frac{\left(8 \hat{\kappa} ^2 H^2-1\right) \log{p_{v}} H'}{2 \hat{\kappa} ^2 H^3}+\frac{2 H'}{H}\right)-\frac{\left(2 G H \hat{\kappa} ^2+G' H'\right)
   \pp p_{v} \pp p_{z}}{4 \hat{\kappa} ^3 G^3 H} \notag \\
&&
+\left(\frac{\pp p_{v}}{2 \hat{\kappa}  G^2}-\frac{\pp^{2} p_{\phi}}{4 \hat{\kappa} ^2 G^4}\right)
   \pz p_{\phi}-\frac{\pp p_{\phi} \pz\pp p_{\phi}}{2 \hat{\kappa} ^2 G^4}+\left(\frac{\pz p_{\phi}}{2 \hat{\kappa} 
   G^2}-\frac{\pp p_{z}}{\hat{\kappa}  G^2}\right) \pv p_{\phi}-\frac{\pp p_{\phi} \pv p_{z}}{\hat{\kappa}  G^2} \notag \\
&&
+\log{p_{v}} \Biggl(\frac{3
   G' (\pp p_{\phi})^2}{2 \hat{\kappa} ^2 G^5}-\frac{3 \pp^{2} p_{z} \pp p_{\phi}}{4 \hat{\kappa} ^2 G^4}-\frac{H' \pz p_{z}
   \pp p_{\phi}}{4 \hat{\kappa} ^2 G^2 H}+\frac{\pv p_{z} \pp p_{\phi}}{2 \hat{\kappa}  G^2}-\frac{H' (\pp p_{z})^2}{2 \hat{\kappa} ^2 G^2
   H}-\frac{3 \pp p_{z} \pp^{2} p_{\phi}}{4 \hat{\kappa} ^2 G^4} \notag \\
&&
-\frac{H' \pp p_{z} \pz p_{\phi}}{4 \hat{\kappa} ^2 G^2
   H}+\frac{\pp p_{z} \pv p_{\phi}}{2 \hat{\kappa}  G^2}\Biggr)+p_{z} \Biggl(\frac{\left(8 G-\frac{G' H'}{\hat{\kappa} ^2 H}\right)
   \pp p_{\phi}}{4 G^3}+2 \pz p_{z}-\frac{2 H' \pv p_{z}}{\hat{\kappa}  H}-\frac{\pv\pp p_{\phi}}{\hat{\kappa}  G^2} \notag \\
&&
+\log{p_{v}}
   \Biggl(\frac{\left(-8 \hat{\kappa} ^2 G H^2+6 G' H' H+G\right) \pp p_{\phi}}{4 \hat{\kappa} ^2 G^3 H^2}-\frac{H' \pp^{2} p_{z}}{2 \hat{\kappa} ^2 G^2
   H}-\frac{\pp^{3} p_{\phi}}{4 \hat{\kappa} ^2 G^4}+\left(2-\frac{1}{2 \hat{\kappa} ^2 H^2}\right) \pz p_{z} \notag \\
&&
-\frac{H' \pz\pp p_{\phi}}{4 \hat{\kappa} ^2 G^2
   H}+\frac{H' \pv p_{z}}{\hat{\kappa}  H}+\frac{\pv\pp p_{\phi}}{2 \hat{\kappa}  G^2}\Biggr)\Biggr).
\end{eqnarray}
%


\end{document}